\documentclass[times,twocolumn,final]{elsarticle}

\usepackage{cag}

\usepackage{amssymb}
\usepackage{amsmath}
\usepackage{multirow}
\newcommand{\revision}[1]{\textcolor{black}{#1}}
\newcommand{\minor}[1]{\textcolor{black}{#1}}
\usepackage{booktabs}
\usepackage[figuresleft]{rotating}
\usepackage{array}
\usepackage{hyperref}
\usepackage{comment}
\usepackage{overpic}

\journal{Compuers \& Graphics}

\begin{document}

\begin{frontmatter}



\title{Point cloud segmentation  for 3D Clothed Human Layering} 


 \author[label1]{Davide Garavaso} 
 \affiliation[label1]{organization={University of Verona},
             addressline={},
              city={Verona},
              postcode={37100},
              country={Italy}}

\author[label1]{Federico Masi} 

 \author[label2]{Pietro Musoni} 
 \affiliation[label2]{organization={University of Parma},
             addressline={},
              city={Parma},
              postcode={43121},
              country={Italy}}

\author[label1]{Umberto Castellani} 


\begin{abstract}
3D Cloth modeling and simulation is essential for avatars creation in several fields, such as fashion, entertainment, and animation. Achieving high-quality results is challenging due to the large variability of clothed body especially in the generation of realistic wrinkles. 3D scan acquisitions provide more accuracy in the representation of real-world objects but lack semantic information that can be inferred with a reliable semantic reconstruction pipeline. To this aim, shape segmentation plays a crucial role in identifying the semantic shape parts. However, current 3D shape segmentation methods are designed for scene understanding and interpretation and only few work is devoted to modeling. In the context of clothed body modeling the segmentation is a preliminary step for fully semantic shape parts reconstruction namely the underlying body and the involved garments. These parts represent several layers with strong overlap in contrast with standard segmentation methods that provide disjoint sets. In this work we propose a new 3D point cloud segmentation paradigm where each 3D point can be simultaneously associated to different layers. In this fashion we can \revision{estimate} the underlying body parts and the unseen clothed regions, i.e., the part of a cloth occluded by the clothed-layer above. We name this segmentation paradigm \textit{clothed human layering}. 
We create a new synthetic dataset that simulates very realistic 3D scans with the ground truth of the involved clothing layers. We propose and evaluate different neural network settings to deal with 3D clothing layering. We considered both coarse and fine grained per-layer garment identification.  Our experiments demonstrates the benefit in introducing proper strategies for the segmentation on the garment domain on both the synthetic and real-world scan datasets. 
\end{abstract}

\begin{graphicalabstract}
\end{graphicalabstract}

\begin{highlights}
\item Research highlight 1
\item Research highlight 2
\end{highlights}


\end{frontmatter}


\section{Introduction}
\label{sec:intro}
\revision{Digital cloth modeling and simulation is a very important task for several applications such as garments design or virtual try-on \cite{vidaurre2020virtualtryon}}. Finding accurate and efficient solutions is challenging due to the wide variability of wrinkles and cloth deformations. Recent approaches exploit data-driven methods based on deep learning and neural network frameworks to improve standard physically-based strategies \cite{Fratarcangeli2018a}. These techniques require the collection of a large amount of reliable annotated data that is not trivial, especially in the 3D domain \cite{musoni2022gim3d,musoni2023gim3dplus}. Early methods rely on synthetic data generated by cloth simulation engines, but their results lack realism \cite{cloth3d}. More recent and promising approaches involve the observation of real examples of clothed humans using 3D scanning techniques \cite{antic2024close,wang20244ddress,zhu2020deep}. These acquisition methods drastically improve the realism but require a suitable processing procedure to infer the reconstruction and semantic interpretation of the raw input data. 
To this aim, 3D cloth segmentation is crucial to identify the underlying body and the involved garments \cite{musoni2022gim3d,musoni2023gim3dplus,antic2024close}. Shape segmentation is a largely studied topic in computer vision and geometry processing \cite{Survey3DSeg}. Best deep-learning-based segmentation results are observed when the 3D data is represented as a 3D point cloud and the neural network involves a transformer-based architecture \cite{zhao2021point,HYY24}. The most investigated applicative context is the segmentation of large 3D scenes composed by several rigid objects like chairs and tables \cite{S3DS}. These scenarios are very different from clothed humans, especially when the overall aim is the semantic reconstruction of body and garments for modeling purposes rather than scene understanding. More precisely, the modeling procedure requires the encoding of body and garments as separated meshes and organized in different overlapping clothing layers \cite{hong2021garmentd}. However, during the scanning phase of real samples only the visible parts are observed even if we know that often this visible part is occluding some underlying garments and body parts\cite{antic2024close,wang20244ddress,zhu2020deep}. Therefore, for the modeling task, it is useful to recover the information of the overlapping layers that cannot be inferred by standard segmentation methods based on the estimation of disjoint sets.          


In this work we propose a new segmentation paradigm for 3D clothed humans. Differently from previous work we estimate the point-label for a multiple set of overlapping clothing layers. The main idea is that a single point can be assigned simultaneously to more semantic parts (i.e., the layers). For instance when a human is dressing a t-shirt above the trousers, a single point around the belt can belong to three layers namely i) body, ii) trousers and iii) t-shirt, even if the visible garment is only the t-shirt.  Therefore, our method consists of estimating a multidimensional label for each point with the dimension given by the number of involved garments plus the body. In this fashion we can also identify the unseen part of a cloth and the underlying part of the body \cite{SMPL}. 
We call this new multidimensional label segmentation approach as \textit{clothed human layering}. 
In order to implement and evaluate this new segmentation technique we developed a new dataset derived from previous work involving several synthetic simulations of clothed humans in different outfits and poses \cite{cloth3d}. 
We use a 3D scanner simulator to generate a realistic 3D cloud point from the synthetic 3D model \cite{koch_piadyk_2021}. 
\revision{This provides us a realistic simulation of 3D scans representing 3D clothed humans with ground truth labels per-point and, unlike with real datasets, also per-layer.}
We propose and evaluate different neural network architectures properly designed to implement this new segmentation paradigm, relying on the new advances in 3D point-cloud processing \cite{pointnet++, dgcnn,zhao2021point}. The key aspect is the definition of a vectorial output for each point rather than the standard scalar labeling. We evaluated different solutions to consider alternative ways to encode the clothing layers and their overlapping areas. 
Our experiments provide promising results for the label estimation of each layer providing a reliable identification of both visible and occluded parts. Moreover, we show interesting qualitative results of clothed human layering on a set of real scan examples \cite{antic2024close}. 

The main contributions of our work are:
\begin{itemize}
	\item We propose the idea of clothed human layering to improve the semantic interpretation and reconstruction of clothed humans observed by a scanning acquisition procedure. 
	\item We provide a new dataset of very realistic simulations of clothed human scans to improve the use of data-driven methods in the cloth modeling and simulation domain. 
	\item We evaluate different strategies to extend the standard 3D point cloud segmentation architecture to work on multidimensional labels.
\end{itemize} 

The rest of the paper is organized as follows. Section \ref{sec:related} revises the state of the art of 3D segmentation on both generic scenes and the specific domain of clothed humans. Section \ref{sec:scanned dataset} introduces our new dataset for clothed human layering describing the 3D scanning simulation procedure and the definition of ground truth annotation.  Section \ref{sec:method} introduces our design of proper neural network architectures and strategies to deal with clothed human layering. Section \ref{sec:results} reports the experimental evaluation for different neural network settings. Finally, Section \ref{sec:conclusions} discusses conclusions and future work.      
\section{Related Works}
\label{sec:related}
Data segmentation is a classic topic in computer vision, geometry processing, and machine learning \cite{duda2000}. Usually, segmentation represents an intermediate step toward high-level data interpretation and understanding. In this section we introduce at first the state of the art on cloth segmentation, then we enlarge the application domain to more generic scenarios for which the 3D data is represented by point cloud.

\paragraph{Clothed-human Datasets} 
One of the main challenges in 3D clothed-human segmentation tasks, particularly in data-driven approaches, is the limited availability of labeled 3D data. Acquiring high-quality segmented data is resource-intensive, requiring specialized equipment for real scans and significant effort for post-processing and manual labeling. These challenges, combined with the growing reliance on deep learning techniques, have led to increasing interest in the development of dedicated datasets for this purpose.
In recent years, several datasets featuring clothed humans have been proposed to address the scarcity of 3D data in clothed human modeling and reasoning. These datasets vary in both characteristics and suitability for segmentation tasks. Some provide real-world data \cite{antic2024close,tiwari20sizer,zhu2020deep, wang20244d}, while others are synthetic \cite{cloth3d, musoni2022gim3d, musoni2023gim3dplus}. \revision{The former capture real-world geometry better but often include noise and they are challenging to label, whereas the latter are easier to process and annotate but typically they face limitations in generalizing to real-world scenarios.} Some datasets provide sequences of real scans \cite{wang20244ddress} which is useful for multi-pose generalization but lack in number of scanned subjects, other datasets like \cite{tiwari20sizer} contain a wide variability in subjects and garments but are limited in a single rest pose. The dataset introduced in \cite{antic2024close} consists of real scans that capture variability in both pose and subject. However, it features repeated instances of the same clothing and poses across multiple scan samples, while also offering a restricted range of poses that excludes more unusual movements. Some works overcome these limitations by introducing synthetic datasets \cite{cloth3d, musoni2022gim3d, musoni2023gim3dplus}. The CLOTH3D dataset \cite{cloth3d} is generated by a wide variety of human animation sequences (including extreme poses), combined with a wide variability of human shapes and garment classes. The works \cite{musoni2022gim3d, musoni2023gim3dplus} expand the concept of CLOTH3D by automatically labeling the dataset and processing the data to simulate the geometry of a real scan. \revision{These datasets, however, are not able to generalize the learning on real scanned subjects enough, in the context of the segmentation tasks. }
Although each of the mentioned datasets has distinct characteristics, a comprehensive dataset that combines accurate real-world geometry representation with extensive pose, subject, and garment diversity — alongside precise labeling for data-driven segmentation — remains absent in the current state-of-the-art.

\paragraph*{Cloth Segmentation}
In the context of clothed humans, the segmentation is very challenging especially when a \emph{per-point-classification} approach is considered, i.e., the point label is chosen among the garment types dressed by the human. Most cloth segmentation methods are proposed for the 2D domain \cite{JSX22,JZX18,SXL15}, where the task is favored by the regular (i.e., euclidean) structure of the image. Moreover, these methods strongly rely on color information that integrates the geometric properties of the shapes. When clothed human segmentation is addressed in the 3D domain the task is still an open issue. The proposed work in this field is very few  
\cite{musoni2022gim3d,musoni2023gim3dplus,antic2024close}. In \cite{musoni2022gim3d,musoni2023gim3dplus} authors propose a new dataset for 3D dressed human segmentation to encourage research in this field. They evaluate the segmentation performances using generic 3D segmentation methods in simplified scenarios where only the upper and lower parts of the body are distinguished. In \cite{antic2024close} more promising results are proposed by introducing a new 3D segmentation method that provides \emph{fine-grained} segmentation results. This method relies not only on the geometric properties of the 3D point cloud but also on color information and the relation between the cloud point and a parametric model of the underlying human, namely SMPL \cite{SMPL}. Other approaches aimed at obtaining a semantic reconstruction of the clothed humans without using an explicit and challenging segmentation procedure \cite{MultiGarmetNet, tiwari20sizer}. These methods adopt a generative approach extending the SMPL parameter model from naked- to dressed- humans where the garment is encoded as a displacement vector from the body. In this fashion, they are more suitable for tight clothes but they are not able to capture the overlapping parts between different clothes. 

\paragraph*{Deep 3D point cloud segmentation}
In the context of the 3D domain, an effective representation of data is the 3D point cloud which is the natural output of the 3D scanning acquisition procedure. Segmentation on point clouds is more challenging than image segmentation due to the absence of a regular structure. In recent years several methods addressed the 3D point segmentation task using deep learning solutions \cite{Survey3DSeg}. The most investigated task is semantic part segmentation or instance object detection on large scale scenarios for which some important benchmarks are publicly available to evaluate the segmentation performance \cite{S3DS,uy-scanobjectnn-iccv19,ShapePart}. 
An early milestone work was proposed in \cite{pointnet} where the PointNet neural network architecture was introduced to effectively combine local information with the global context. This method was extended to deal with hierarchical structure \cite{pointnet++}, and to improve the training strategy \cite{pointnext}. More recently, the 3D point segmentation has been addressed using Point Transformer architectures \cite{zhao2021point}. The first version of this network was extended to improve the speed \cite{wu2022point,park2022fast} or to better encode the relation between local and global information \cite{HYY24}. 
Recent works, such as \cite{openvocabulary2025}, combine 3D models and textual information for semantic part segmentation of clothed humans, leveraging the rapid advancement of large language models (LLMs) to tackle complex tasks such as 3D human segmentation.


         
%
\revision{
\paragraph*{Our work}
 In this work, we introduce the idea of clothed human layering to make the semantic modeling reconstruction easier.
 }
 Unlike previous methods, we modify the structure of state-of-the-art \revision{segmentation} architectures to simultaneously predict both visible and hidden clothing layers within a single scan-like point cloud. 
 \revision{This is a peculiar characteristic of our segmentation framework that has not been investigated in the literature yet.}
 To support this approach, we developed a dedicated synthetic dataset that accurately replicates the characteristics of real-world scans using a digital scanner simulator applied to synthetic data. 












\section{Scanned dataset}
\label{sec:scanned dataset}

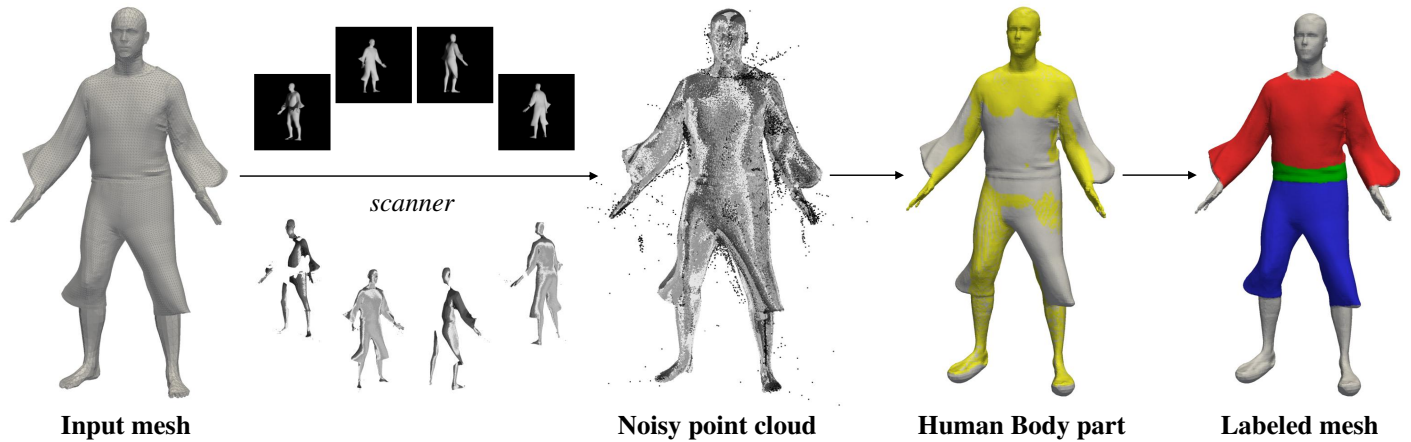
\begin{figure*}[!t]
    \centering
    \begin{overpic}[
        trim=0cm 0cm 0cm 0cm,
        clip,
        width=\linewidth
    ]{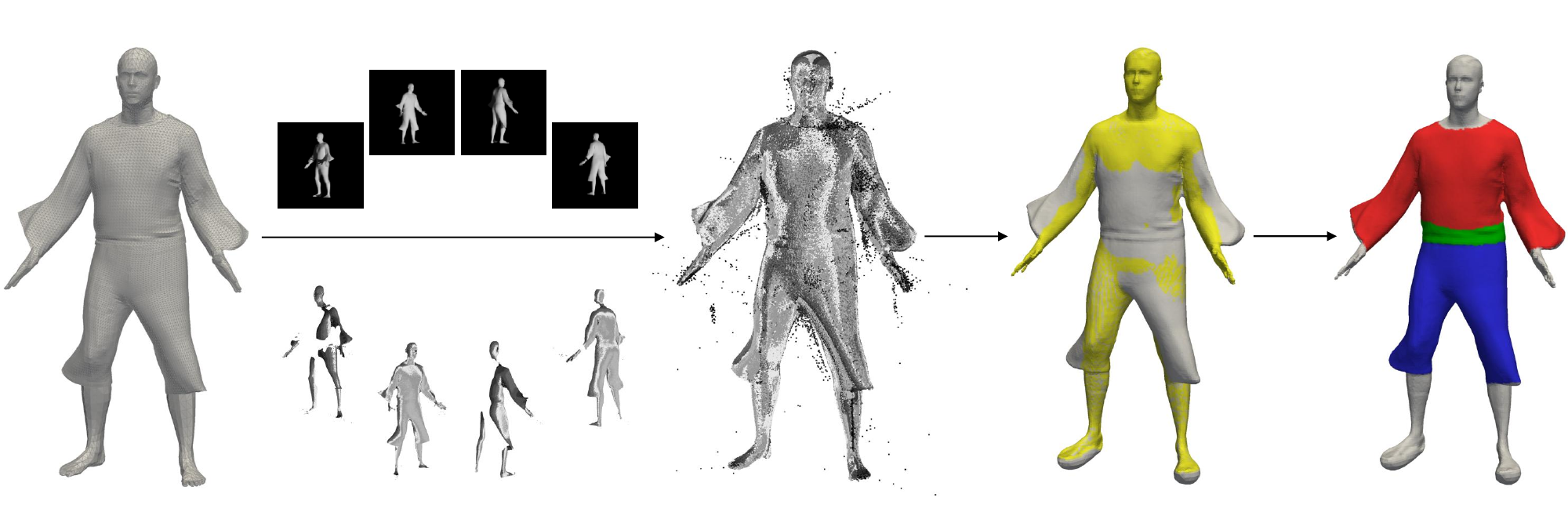}
        \put(26, 15) {\textit{scanner}}
        \put(3.4,-1){ \textbf{Input mesh}}
        \put(43.3,-1){ \textbf{Noisy point cloud}}
        \put(64.8,-1){ \textbf{Human Body part}}
        \put(86.5,-1){ \textbf{Labeled mesh}}
    \end{overpic}
    \caption{
        Here is the pipeline for constructing the scanned dataset. From left to right: the input mesh to the structured light scanner simulator, which produces a scan-like point cloud with noise; the estimated body points (highlighted in yellow) on the point cloud; and finally, the labeled mesh, where red points represent the T-shirt, blue points represent the pants, and green points indicate the overlap regions.
    }
    \label{fig:scanned_dataset}
\end{figure*}

\begin{figure}[th]
    \centering
    \includegraphics[width=\linewidth]{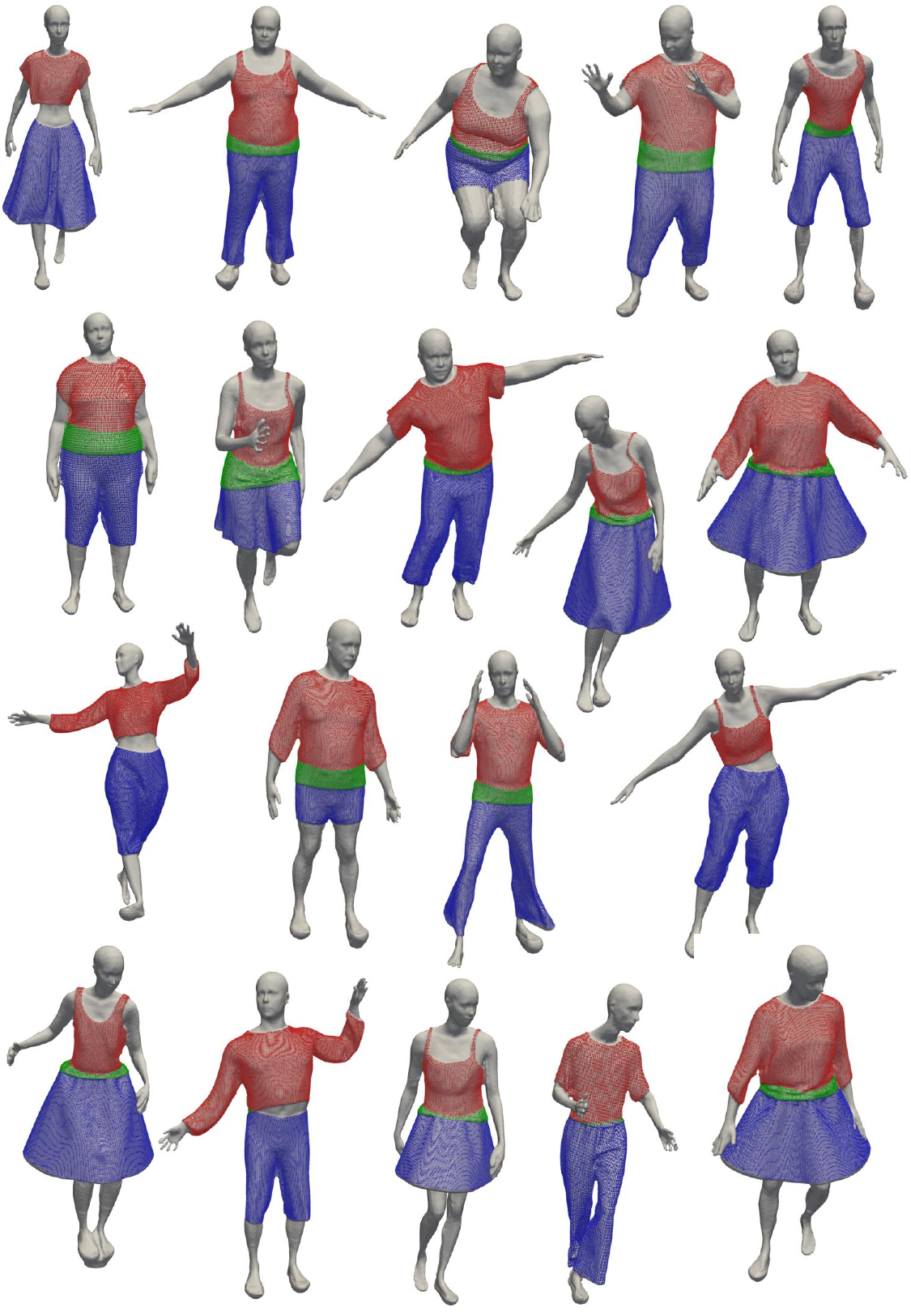}
    \caption{
        \revision{Here some examples from our dataset of human meshes, including both male and female subjects with a wide range of body shapes and clothing types, with and without cloth overlap.}
    }
    \label{fig:scanned dataset}
\end{figure}

Obtaining real 3D scans of clothed human subjects is both expensive and logistically challenging. In addition, even high-quality scans fail to capture hidden parts of garments, such as overlapping or occluded clothing layers. We addressed these limitations by introducing a synthetic scan dataset composed of 3D point clouds comparable to real scans including multilayer information, explicitly revealing overlapping regions of worn garments. We started from individual synthetic garment meshes provided by the CLOTH3D dataset \cite{cloth3d}, specifically selected within the GIM3D plus dataset\cite{musoni2023gim3dplus}, and merged them into complete human outfits. Moreover, to increase the size of the dataset, we generated additional samples by extracting garment meshes from multiple animation frames, for a maximum of 3 frames for each subject. \\
The resulting clothed human meshes were then scanned using a structured light scanner simulator \cite{koch_piadyk_2021} that produces scan-like point clouds. For our dataset, each mesh was scanned from 13 different viewpoints, selected to avoid capturing the underside of the feet, mimicking the limitations of real-world scanning setups. The generated point clouds exhibit realistic noise and a spatial distribution similar to actual structured light scans, maintaining consistent vertex density across samples regardless of mesh topologies, which facilitates reliable reconstruction and segmentation. After the reconstitution step, we were able to label the garment regions on the scanned mesh by projecting information from the original clothes. The key strength of this dataset lies on the fact that we have access to the original separate garments, letting us accurately identify the overlapping areas with the knowledge of both the visible and not visible garment parts; e.g., when a t-shirt extends over a pair of trousers, the overlapping area is explicitly labeled as belonging to both garments even though the underlying trouser region is not visible. The same process was applied to the human body: starting from the SMPL \cite{SMPL} vertices, the closest points on the scanned mesh were identified, providing a reconstruction of the human body for both the visible part and the part underneath the clothing. In practice,  the under-the-cloth body is detected as the tightest garment parts.   
Our scanned data set is composed of 3306 labeled meshes with an average density of about 50000 vertices, in Table \ref{tab:scan dataset} the subdivision of the data is shown in all clothing categories. 
\revision{
Several samples of our datasets are shown in Figure \ref{fig:scanned dataset}. The dataset is public available here \footnote{Removed for blind review.}.
}

\begin{table}[t]
    \centering
    \begin{tabular}{l c c c c}
    \toprule
    Clothing category & Fr0 & Fr100 & Fr200 & Total \\
    \cmidrule(lr){1-1}
        \cmidrule(lr){2-4}
        \cmidrule(lr){5-5}

    long-shirt + skirt & 73 & 37 & 24 & 140 \\
    long-shirt + trousers & 151 & 88 & 69 & 308 \\
    long-shirt + shorts & 162 & 142 & 84 & 388 \\
    t-shirt + skirt & 98 & 29 & 27 & 154 \\
    t-shirt + trousers & 257 & 142 & 105 & 504 \\
    t-shirt + shorts & 243 & 151 & 106 & 500 \\
    top + skirt & 161 & 104 & 45 & 310 \\
    top + trousers & 426 & 74 & 0 & 500 \\
    top + shorts & 340 & 162 & 0 & 502 \\
    \bottomrule    
    \end{tabular}
    \label{tab:scan dataset}
    \caption{Here is the composition of the dataset divided by clothing categories. the first column lists the outfit combinations, while the next three columns show the number of meshes scanned at different frames of the same animation. The last column reports the total number for each outfit. The last two rows do not include samples at frame 200, as we aimed to balance the dataset and avoid an overrepresentation of specific garments. The final dataset consists of 3306 mesh, including 836 long-shirts, 1158 t-shirts, 1312 tops, 1312 trousers, 1390 shorts and 604 skirts.}
\end{table}

\section{Method}
\label{sec:method}
In this section, we introduce the most promising deep-learning-based segmentation methods that have been proposed for generic 3D point cloud segmentation and, in particular, we use them as feature extractors. Then, we propose different strategies to extend these neural network architectures to deal with multiple \minor{clothing} layers. We highlight how to settle the neural network parameters to obtain coarse segmentation (i.e., lower and upper garments segmentation) or fine-grained segmentation.

\minor{
For clarity, we note that throughout this chapter, the term ``layer'' specifically refers to a single component of the network’s vectorial output, unless otherwise stated. This usage is distinct from both its broader meaning as an internal layer in neural network architectures and the ``clothing layer'' defined earlier in the text.
}

\subsection{Feature extraction}
In order to extract the features that will compose the backbone of the segmentation architecture we exploit PointNet++~\cite{pointnet++}, Dynamic Graph CNN~\cite{dgcnn}, and Point Transformer~\cite{zhao2021point}. 

\paragraph{PointNet++}

The PointNet++ architecture~\cite{pointnet++} is based on the PointNet neural network~\cite{pointnet} that was a pioneer work to study deep learning on point sets. The PointNet architecture is based on the estimation of a global feature vector that encodes the overall properties of the shape. Then, this vector is concatenated to each point to combine local and global information. PointNet++ aims at exploiting the local information in a hierarchical fashion. It applies PointNet recursively on a nested partitioning of the input cloud point. Therefore, PointNet++is able to learn local features with increasing contextual scales. 

\paragraph{Dynamic Graph CNN}

The Dynamic Graph Convolution Neural Network~\cite{dgcnn}~\minor{(DGCNN)} exploits the neighborhood of a point to encode local and global information. The key component is the edge convolution block that extends the standard convolution operator from images to direct graphs. Rather than computing a fixed graph for the entire point cloud, the graph is dynamically computed at each point for each level. Several \minor{network layers} are defined where the k-nearest neighbor is computed to identify the local context in the feature space. In this fashion, the network learns how to construct the graph instead of relying on a predefined structure. The feature vectors extracted at each level are concatenated to get the output.

\paragraph{PointTransformer}

The Point Transformer architecture~\cite{zhao2021point} is designed to process point cloud data by leveraging self-attention mechanisms to capture local geometric structures. By applying self-attention on points at different scales, the network learns internal data representations for the segmentation task. For each point in the input point cloud, the point transformer block applies self-attention to a local patch of the k-nearest neighbor for the point. Positional encoding is embedded as relative coordinates. The block enhances feature representation without altering the feature dimensionality or the number of points. The point transformer network is composed of an encoder that employs a set of transition down steps to explore the point cloud in a hierarchical fashion. The encoder is followed by a decoder that implements the transition up steps to recover the same number of input points.   

\minor{
\paragraph{Architectural selection rationale}
The three models presented were chosen as seminal contributions that have shaped the development of deep learning techniques for processing raw point sets.
Each introduces a foundational paradigm—hierarchical feature learning, edge convolutions, and attention mechanisms—that continues to influence subsequent research.
Approaches based on voxelization and sparse convolutions were excluded, as they rely on fundamentally different representations.
Preference was given to architectures with proven reliability and conceptual clarity, prioritizing robustness and interpretability over the increased complexity often introduced by more recent methods.
}

\subsection{Extension to multi-layers}
The peculiarity of our task consists of dealing with 3D clothed humans that require a different segmentation paradigm than other application domains.
According to our idea of clothed human layering, we propose a new segmentation framework that allows the assignment of multiple labels to the same point.
\minor{A layer thus is defined by the segmentation of a single label over the whole point cloud.  We define strategies that differ on the number of layers involved and the semantics associated to each layer.}

\minor{Given a strategy,} the simplest and trivial solution is \minor{to train} an independent neural network for each layer. 
This is the most computationally expensive approach that totally ignores the correlation between the layers.
A more interesting solution is instead \minor{to use} a unified neural network that learns how to select the useful features to estimate the correct labels for each layer.
Similarly to standard neural network architecture\minor{s} for point cloud segmentation, we propose to implement the segmentation as a per-point classification task.
The core idea is to estimate a single feature vector for each point and feed it to different multilayer perceptrons (MLPs), each one associated to a different layer.

\begin{figure}[th]
    \centering
    \includegraphics[width=\linewidth]{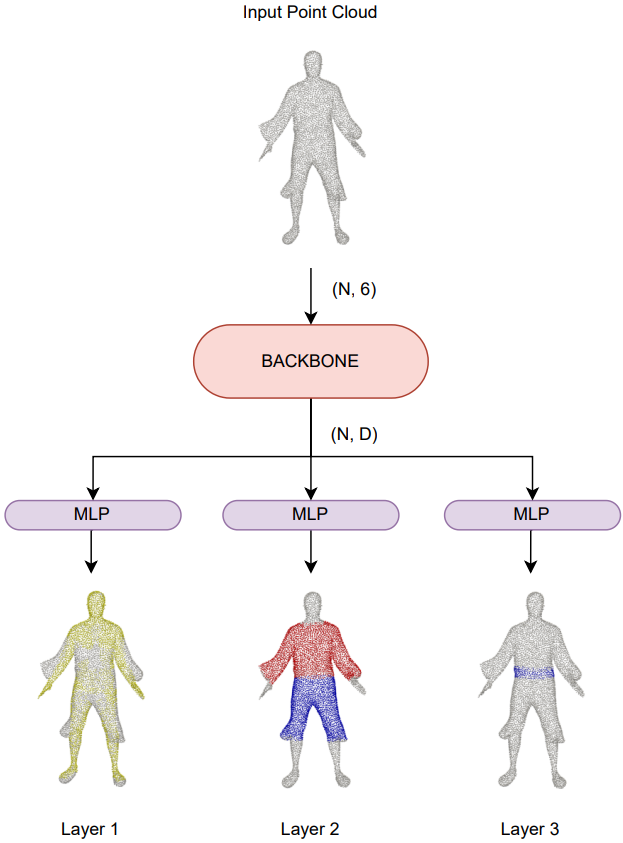}
    \caption{
        Multilayer architecture.
        The backbone receives as input the coordinates and normals of a point cloud, and produces per each point an embedding feature vector.
        From that, multiple MLPs are used to generate the segmentation for each layer. In this figure the point cloud at each layer are colored according to Strategy 3 (see below).
    }
    \label{fig:multihead-arch}
\end{figure}

Figure \ref{fig:multihead-arch} shows a scheme of the proposed neural architecture. The input point cloud is processed by the backbone component that estimates a feature vector for each 3D point. As described in the previous section the backbone can be chosen between i) PointNet++, ii) DGCNN, or iii) PointTransformer.  Then, each MLP provides the pointwise label for each layer.     

In the following, \revision{we design different strategies, based on the proposed architecture, to solve the clothed human layering}. The first three strategies address the coarse cloth segmentation task that provides a distinction between upper- and lower- garment parts. The last two strategies introduce a fine-grained segmentation procedure. These strategies differ also on the way we encode the overlapping between the layers, i.e., implicit and explicit.    

\begin{figure*}[!h]
    \centering
    \begin{overpic}[
        trim=0cm 0cm 0cm 0cm,
        clip,
        width=\linewidth
    ]{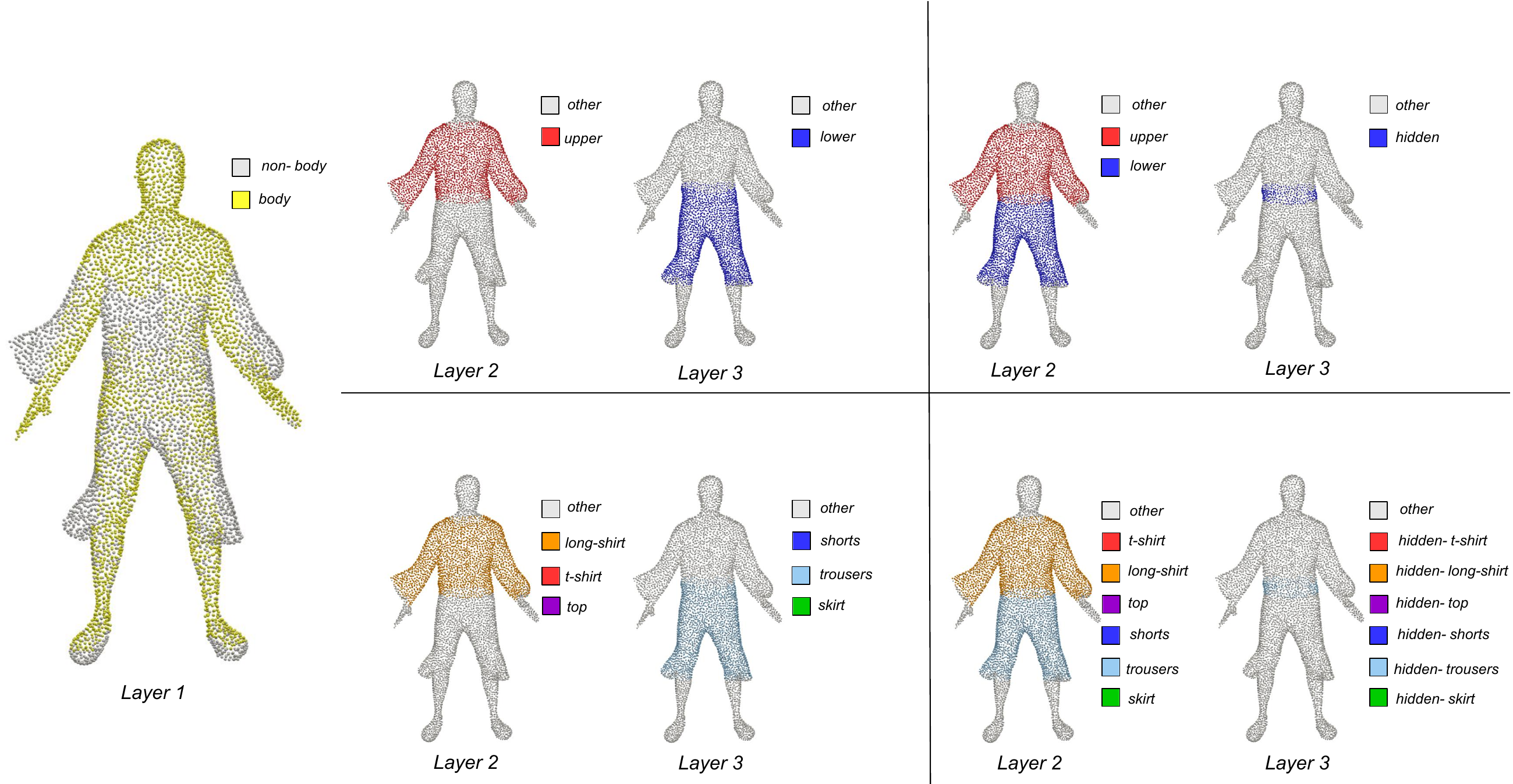}
        \put(36.7, 49.2) {\textbf{Strategy 2}}
        \put(78.2, 49.2) {\textbf{Strategy 3}}
        \put(36.7, 22) {\textbf{Strategy 4}}
        \put(78.2, 22) {\textbf{Strategy 5}}

    \end{overpic}
    \caption{
        Here we show the different approaches we adopt for multilayer segmentation. The first layer, common to all strategies, represents the adherence of clothing to the underlying body and the body visible parts. In Strategy 2, the second layer employs a binary label to indicate the presence of an upper garment, while the third layer identifies the lower garment. Strategy 3 merges the upper and lower garment labels into the second layer and reserves the third layer for the hidden portions of the clothing. Strategies 4 and 5 extend the segmentation task by also predicting specific garment classes. The former assigns distinct labels in layer 2 for upper garments and in layer 3 for lower garments, whereas the latter incorporates all seven garment classes in layer 2 for visible clothing and applies the same classification in layer 3 for hidden clothing. 
    }
    \label{fig:strategies}
\end{figure*}

\paragraph*{\textbf{Strategy 1}: Coarse segmentation with standard architecture}

In this strategy, we implement a standard neural network architecture with a single final MLP devoted to deciding the pointwise label. \minor{This is equivalent to a multilayer strategy with one layer only.} One of the labels is dedicated to the overlapping part. The others encode the upper-garment, the lower-garment, and the rest of the pointcloud (i.e., mainly the visible body-part). In this strategy, the full-body layer is not considered.

\paragraph*{\textbf{Strategy 2}: coarse segmentation with implicit overlap}

In this strategy, we associate an MLP to each layer. The first \minor{layer encodes full-body information} (i.e., a binary label for body or non-body point). The second layer encodes the upper garment (i.e., a binary label for upper garment or non-upper garment point). The third layer encodes the lower garment (i.e., a binary label for lower-garment or non-lower garment point). Note that the part of garment overlap is encoded implicitly as the only points that belong simultaneously to both upper and lower garment.
From this encoding procedure, the output layers of the neural network are ready to be directly converted into three independent meshes for modeling purposes.   
%

\paragraph*{\textbf{Strategy 3}: coarse segmentation with explicit overlap}

In this strategy, the first layer is the same as the previous strategy. The second layer provides the segmentation of the visible part, like in standard segmentation procedures. The third layer is  dedicated to the hidden clothed parts.
\minor{Here, we explicitly detect overlapping garment regions in the point cloud. This strategy allows us to infer the visible layering of garments directly from the data. In our dataset—as well as in all real scans from the datasets we tested—the upper garment overlaps the lower one. For example, in a t-shirt–skirt outfit, we can infer that the hidden part corresponds to the skirt, while the t-shirt is the visible garment.}
%

\paragraph*{\textbf{Strategy 4}: \revision{fine-grained} segmentation with implicit overlap}

This strategy extends Strategy 2 to fine-grained segmentation. \minor{The second layer still encodes the upper garments but with the label assuming values among the possible garment types of the upper body part. Similarly, the third layer encodes lower-body garments using the same principle.} Note that in this strategy we can exploit the fact that only a subset of garment types can be associated with the upper body part. For instance, a skirt cannot be assigned to the upper garment and therefore it will be not considered as a valid value for the second layer (\minor{an analogous reasoning applies for the third layer}).

\paragraph*{\textbf{Strategy 5}: \revision{fine-grained} segmentation with explicit overlap}

In this strategy, we extend Strategy 3 to fine-grained segmentation. \minor{The second layer} encodes the fine-grained segmentation of the visible part. Therefore, the label value \minor{can correspond} to all the involved garment types. \minor{Analogously, the third layer} encodes the fine-grained segmentation of the \minor{hidden} garment part. Note that in this strategy we cannot reduce the number of garment labels per layer since in principle both the visible and hidden layer can involve any garment type. \minor{This strategy explicitly encodes which garment is worn underneath (and thus not visible) and which is worn on top. This makes the task more challenging but also more suitable for upgrades to new configurations, such as introducing new combinations in our dataset—for instance, when the lower garment is visible (e.g., shorts) while the upper garment is not (e.g., a t-shirt).}

Figure \ref{fig:strategies} reports a visual representation of the labeling procedure for strategies 2-5.


\section{Results}
\label{sec:results}

In this section, we report an extensive evaluation of the proposed 3D clothed human layering task. We clarify implementation details and the validation protocol. Then, we introduce the quantitative results for each proposed strategy on the realistic synthetic dataset described in Section~\ref{sec:scanned dataset}. Finally, additional qualitative results are shown \revision{on real data} to verify the capability of the neural network  \revision{trained on synthetic data in generalizing the segmentation on} previously unseen 3D real scan examples. 

\subsection{Implementation Details}

All models were implemented using the Pointcept framework~\cite{pointcept2023}, which provides efficient CUDA-accelerated point cloud operations. The existing Point Transformer implementation was extended to support the aforementioned strategies. To ensure consistency across experiments, PointNet++ and DGCNN were re-implemented within the same framework. The training was conducted using the AdamW optimizer with a one-cycle learning rate policy employing cosine annealing and a peak learning rate of 0.005. Data augmentation, when applied, included random rotation, scaling, and translation of point clouds. All experiments were executed on an RTX 4090 GPU using float16 precision and batch sizes maximized to utilize the full 24 GB of available VRAM.

\subsection{Validation protocol}

The segmentation task is framed as a per-point classification problem, with performance evaluated using the Intersection over Union (IoU) metric. The mean IoU (mIoU) is computed for each architectural layer as the average IoU across all classes present in that layer. Subsequently, the final mIoU is obtained by averaging the layer-wise mIoUs. For each class, the IoU is defined as the ratio of the number of points in the whole dataset where both the prediction and ground truth share the class label to the number of points where either the prediction or the ground truth assigns that label; equivalently, IoU can be expressed as

\begin{equation}
    \frac{\text{TP}}{\text{TP} + \text{FN} + \text{FP}},
\end{equation}

where TP, FN, and FP denote true positives, false negatives, and false positives, respectively. Evaluation results are reported in Tables~\ref{tab:strategy1-results}--\ref{tab:strategy5-results}.
\revision{In addition to IoU and mIoU, Table~\ref{tab:strategy1-results} reports mean accuracy (mAcc), computed as the average per-class accuracy where only points of the given class are considered, and overall accuracy (allAcc), which reflects the proportion of correctly classified points disregarding of  class information.  For the other strategies, only mIoU and IoU are reported to maintain a clear and compact table layout.}

\subsection{Experimental evaluation}
The experimental evaluation is carried out on i) coarse segmentation using a single layer, ii) coarse segmentation with multiple layers, and iii) fine-grained segmentation with multiple layers.

\paragraph*{Coarse segmentation using a single layer}

This experiment implements Strategy 1 described in Section \ref{sec:method}. The neural architecture is the standard one for 3D point cloud segmentation. We introduce an additional label to encode the overlapping part between the upper and the lower garments. The evaluation of this additional label is the most interesting to appreciate the 3D clothed human layering task. Results are reported in Table \ref{tab:strategy1-results} and 
 are evaluated for the three feature extractors described in Section \ref{sec:method}. Results are satisfying, especially when the Transformer architecture is considered and combined with data augmentation. 
 This strategy is limited since the body layer is not considered. \revision{The overlap label is correctly detected (Table \ref{tab:strategy1-results} shows $79.4\%$ per-class IoU) but it performed worse than other classes.}
 The variability of chosen labels is reduced due to the coarse segmentation task. The extension of this strategy to fine-grained segmentation considering the layering idea is complicated since it will require an explosion of involved labels.  

\paragraph*{Coarse segmentation with multiple layer}

This experiment implements Strategy 2 and 3 and provides results shown in Table\ref{tab:strategy2-results} and Table \ref{tab:strategy3-results}  respectively. Strategy 2 implements a direct relation between the final layers of the network and the semantic clothed layers, as expected by 3D artists where the underlying body and the involved garments are encoded in separated meshes. The first layer that encodes the body part is the most challenging since it contains both visible and hidden areas. Results are satisfying even if there is a large margin of improvement. This layer contains an unbalanced set of points toward the 'body' label and therefore a proper strategy should be designed to consider this aspect. The second and third layers instead are well-balanced and provide promising results. The limitation of this strategy is the lack of information about the visible and hidden parts of the overlapping garments. Strategy 3 implements an explicit encoding of the hidden garment part, and the behavior of Layer 1 is very similar to Strategy 2. Layer 2 implements a segmentation of visible parts as standard 3D segmentation methods. It is interesting that performances shown in this layer are very promising reaching a mean IOU of $97.4\%$. Layer 3 suffers by the similar limitation of data unbalancing toward non-hidden labels. Also for this layer, some specific strategies can be exploited to improve the performance.          
  
\begin{table*}[tbp]
    \centering
    \begin{tabular}{l *{3}{c} *{4}{c}}
        \toprule
            Method & mIoU & mAcc & allAcc & other & upper & overlap & lower \\
        \cmidrule(lr){1-1}
        \cmidrule(lr){2-4}
        \cmidrule(lr){5-8}
            DGCNN                          &          $79.5$ &          $88.0$ &          $92.2$ &          $85.9$ &          $82.9$ &          $59.6$ &          $89.6$ \\
            DGCNN aug                      &          $76.2$ &          $85.2$ &          $91.2$ &          $84.6$ &          $81.2$ &          $50.5$ &          $88.6$ \\
            PointNet++                     &          $79.2$ &          $88.6$ &          $93.1$ &          $88.8$ &          $84.7$ &          $52.1$ &          $91.1$ \\
            Pointnet++ aug                 &          $82.1$ &          $89.0$ &          $94.4$ &          $90.8$ &          $86.9$ &          $57.6$ &          $93.0$ \\
            PointTransformer v1            &          $91.7$ &          $95.3$ &          $97.7$ &          $96.3$ &          $93.6$ &          $79.1$ &          $97.6$ \\
            PointTransformer v1 aug        & \boldmath$92.1$ & \boldmath$95.6$ & \boldmath$98.0$ & \boldmath$96.9$ & \boldmath$94.2$ & \boldmath$79.4$ & \boldmath$98.0$ \\
        \bottomrule
    \end{tabular}
    
    \caption{
        \revision{Single layer strategy \minor{(Strategy 1)} quantitative results: mean IoU (mIoU), mean accuracy (mAcc), overall accuracy (allAcc), and per-class IoU scores}.
    }\label{tab:strategy1-results}
\end{table*}

\begin{table*}[tbp]
    \begin{tabular}{l c *{3}{c} *{3}{c} *{3}{c}}
        \toprule
            \multirow{2}{*}{Backbone} & \multirow{2}{*}{avg mIoU} & \multicolumn{3}{c}{Layer 1} & \multicolumn{3}{c}{Layer 2} & \multicolumn{3}{c}{Layer 3} \\
                                      &                           & mIoU & no-body & body       & mIoU & other & upper        & mIoU & other & lower        \\
        \cmidrule(lr){1-1}
        \cmidrule(lr){2-2}
        \cmidrule(lr){3-5}
        \cmidrule(lr){6-8}
        \cmidrule(lr){9-11}
            DGCNN                          &          $82.8$ &          $64.7$ &          $56.5$ &          $72.9$ &          $89.9$ &          $94.4$ &          $85.5$ &          $93.7$ &          $95.4$ &          $92.0$ \\
            DGCNN aug                      &          $81.3$ &          $62.7$ &          $54.8$ &          $70.6$ &          $88.0$ &          $93.2$ &          $82.7$ &          $93.3$ &          $95.0$ &          $91.5$ \\
            PointNet++                     &          $84.1$ &          $66.0$ &          $58.7$ &          $73.3$ &          $91.5$ &          $95.3$ &          $87.8$ &          $94.6$ &          $96.1$ &          $93.2$ \\
            PointNet++ aug                 &          $85.5$ &          $67.1$ &          $59.9$ &          $74.3$ &          $93.7$ &          $96.5$ &          $90.9$ &          $95.8$ &          $96.9$ &          $94.6$ \\
            PointTransformer v1            &          $88.3$ &          $70.4$ &          $63.4$ &          $77.3$ &          $97.1$ &          $98.4$ &          $95.8$ &          $97.5$ &          $98.2$ &          $96.8$ \\
            PointTransformer v1 aug        & \boldmath$89.0$ & \boldmath$71.5$ & \boldmath$65.0$ & \boldmath$78.1$ & \boldmath$97.8$ & \boldmath$98.8$ & \boldmath$96.7$ & \boldmath$97.9$ & \boldmath$98.5$ & \boldmath$97.3$ \\
        \bottomrule
    \end{tabular}
    \caption{\revision{Strategy 2 quantitative results: average mean IoU across all layers (avg mIoU), then for each layer mean IoU (mIoU) and per-class IoU scores}.}
    \label{tab:strategy2-results}
\end{table*}

\begin{table*}[tbp]
    \resizebox{\textwidth}{!}{
        \begin{tabular}{l c *{3}{c} *{4}{c} *{3}{c}}
            \toprule
                \multirow{2}{*}{Backbone} & \multirow{2}{*}{avg mIoU} & \multicolumn{3}{c}{Layer 1} & \multicolumn{4}{c}{Layer 2}     & \multicolumn{3}{c}{Layer 3} \\
                                          &                           & mIoU & no-body & body       & mIoU & other & upper & lower    & mIoU & other & hidden       \\
            \cmidrule(lr){1-1}
            \cmidrule(lr){2-2}
            \cmidrule(lr){3-5}
            \cmidrule(lr){6-9}
            \cmidrule(lr){10-12}
                DGCNN                          &          $77.8$ &          $64.6$ &          $56.3$ &          $72.9$ &          $87.4$ &          $86.9$ &          $84.8$ &          $90.6$ &          $81.2$ &          $98.7$ &          $63.8$ \\
                DGCNN aug                      &          $75.4$ &          $62.6$ &          $54.2$ &          $70.9$ &          $85.7$ &          $85.1$ &          $82.3$ &          $89.7$ &          $77.9$ &          $98.4$ &          $57.4$ \\
                PointNet++                     &          $78.7$ &          $65.9$ &          $58.2$ &          $73.6$ &          $90.9$ &          $91.1$ &          $88.4$ &          $93.1$ &          $79.2$ &          $98.4$ &          $60.1$ \\
                PointNet++ aug                 &          $80.4$ &          $66.7$ &          $59.3$ &          $74.1$ &          $92.6$ &          $92.6$ &          $90.7$ &          $94.6$ &          $81.8$ &          $98.7$ &          $64.9$ \\
                PointTransformer v1            &          $84.6$ &          $70.0$ &          $63.0$ &          $77.0$ &          $95.7$ &          $95.6$ &          $94.8$ &          $96.8$ &          $88.2$ &          $99.2$ &          $77.1$ \\
                PointTransformer v1 aug        & \boldmath$86.2$ & \boldmath$71.5$ & \boldmath$64.8$ & \boldmath$78.2$ & \boldmath$97.4$ & \boldmath$97.2$ & \boldmath$96.7$ & \boldmath$98.3$ & \boldmath$89.9$ & \boldmath$99.3$ & \boldmath$80.4$ \\         
            \bottomrule
        \end{tabular}
    }
    \caption{\revision{Strategy 3 quantitative results: average mean IoU across all layers (avg mIoU), then for each layer mean IoU (mIoU) and per-class IoU scores}.}\label{tab:strategy3-results}
\end{table*}

\begin{table*}[tbp]
    \resizebox{\textwidth}{!}{
        \begin{tabular}{l c *{3}{c} *{5}{c} *{5}{c}}
            \toprule
                \multirow{2}{*}{Backbone} & \multirow{2}{*}{avg mIoU} & \multicolumn{3}{c}{Layer 1} & \multicolumn{5}{c}{Layer 2}               & \multicolumn{5}{c}{Layer 3}                \\
                                          &                           & mIoU & no-body & body       & mIoU & other & long-shirt & t-shirt & top & mIoU & other & long-pants & shorts & skirt \\
            \cmidrule(lr){1-1}
            \cmidrule(lr){2-2}
            \cmidrule(lr){3-5}
            \cmidrule(lr){6-10}
            \cmidrule(lr){11-15}
                DGCNN                          &          $71.3$ &          $63.4$ &          $55.3$ &          $71.5$ &          $64.2$ &          $93.1$ &          $47.5$ &          $47.8$ &          $68.3$ &          $86.3$ &          $94.6$ &          $83.4$ &          $78.6$ &          $88.6$ \\
                DGCNN aug                      &          $71.0$ &          $61.1$ &          $52.5$ &          $69.7$ &          $66.0$ &          $92.0$ &          $57.6$ &          $53.0$ &          $61.6$ &          $86.0$ &          $94.2$ &          $84.3$ &          $77.3$ &          $88.4$ \\
                PointNet++                     &          $75.3$ &          $64.8$ &          $57.6$ &          $72.0$ &          $73.3$ &          $94.8$ &          $63.5$ &          $59.0$ &          $75.9$ &          $87.8$ &          $95.8$ &          $86.2$ &          $77.3$ &          $92.0$ \\
                PointNet++ aug                 &          $75.6$ &          $63.8$ &          $55.8$ &          $71.8$ &          $74.2$ &          $95.0$ &          $64.6$ &          $60.5$ &          $76.7$ &          $88.7$ &          $95.9$ &          $86.9$ &          $79.5$ &          $92.6$ \\
                PointTransformer v1            &          $82.2$ &          $68.3$ &          $60.8$ &          $75.8$ &          $85.3$ &          $97.5$ & \boldmath$78.2$ & \boldmath$77.2$ &          $88.4$ &          $92.8$ &          $97.7$ &          $91.1$ &          $86.0$ &          $96.5$ \\
                PointTransformer v1 aug        & \boldmath$82.8$ & \boldmath$69.1$ & \boldmath$62.1$ & \boldmath$76.1$ & \boldmath$85.6$ & \boldmath$98.5$ &          $76.5$ &          $75.0$ & \boldmath$92.3$ & \boldmath$93.8$ & \boldmath$98.3$ & \boldmath$91.8$ & \boldmath$87.9$ & \boldmath$97.4$ \\             
            \bottomrule
        \end{tabular}
    }
    \caption{\revision{Strategy 4 quantitative results: average mean IoU across all layers (avg mIoU), then for each layer mean IoU (mIoU) and per-class IoU scores}.}\label{tab:strategy4-results}
\end{table*}

\begin{table*}[tbp]
    \resizebox{\textwidth}{!}{
        \begin{tabular}{l c *{3}{c} *{8}{c} *{5}{c}}
            \toprule
                \multirow{2}{*}{Backbone} & \multirow{2}{*}{avg mIoU} & \multicolumn{3}{c}{Layer 1} & \multicolumn{8}{c}{Layer 2}                                             & \multicolumn{5}{c}{Layer 3}                \\
                                          &                           & mIoU & no-body & body       & mIoU & other & t-shirt & shorts & long-pants & long-shirt & top & skirt & mIoU & other & skirt & shorts & long-pants \\
            \cmidrule(lr){1-1}
            \cmidrule(lr){2-2}
            \cmidrule(lr){3-5}
            \cmidrule(lr){6-13}
            \cmidrule(lr){14-18}
                DGCNN                          &          $65.7$ &          $63.3$ &          $54.4$ &          $72.1$ &          $69.9$ &          $84.2$ &          $49.0$ &          $73.3$ &          $81.0$ &          $46.0$ &          $69.3$ &          $86.3$ &          $63.9$ &          $98.4$ &          $50.0$ &          $52.2$ &          $54.8$ \\
                DGCNN aug                      &          $63.5$ &          $61.4$ &          $53.4$ &          $69.4$ &          $70.1$ &          $82.5$ &          $51.8$ &          $71.0$ &          $80.3$ &          $59.7$ &          $59.5$ &          $86.0$ &          $59.0$ &          $97.9$ &          $45.2$ &          $47.9$ &          $45.1$ \\
                PointNet++                     &          $69.3$ &          $65.1$ &          $57.3$ &          $72.9$ &          $79.5$ &          $90.7$ &          $61.6$ &          $80.0$ &          $88.8$ &          $64.0$ &          $78.5$ &          $92.6$ &          $63.5$ &          $98.3$ &          $47.8$ &          $50.4$ &          $57.4$ \\
                PointNet++ aug                 &          $68.5$ &          $64.3$ &          $56.8$ &          $71.9$ &          $79.1$ &          $90.5$ &          $61.8$ &          $79.2$ &          $88.0$ &          $65.7$ &          $77.3$ &          $91.2$ &          $61.9$ &          $98.2$ &          $46.7$ &          $49.5$ &          $53.3$ \\
                PointTransformer v1            &          $77.7$ &          $68.0$ &          $60.5$ &          $75.5$ &          $86.7$ &          $95.1$ &          $72.2$ &          $87.7$ &          $91.9$ &          $76.8$ &          $86.6$ &          $96.7$ &          $78.5$ &          $99.1$ &          $70.7$ &          $71.7$ &          $72.6$ \\
                PointTransformer v1 aug        & \boldmath$79.0$ & \boldmath$68.8$ & \boldmath$61.8$ & \boldmath$75.8$ & \boldmath$89.0$ & \boldmath$96.3$ & \boldmath$77.2$ & \boldmath$88.4$ & \boldmath$93.0$ & \boldmath$77.1$ & \boldmath$93.0$ & \boldmath$97.7$ & \boldmath$79.2$ & \boldmath$99.1$ & \boldmath$71.9$ & \boldmath$72.3$ & \boldmath$73.5$ \\
            \bottomrule
        \end{tabular}
        }

    \caption{\revision{Strategy 5 quantitative results: average mean IoU across all layers (avg mIoU), then for each layer mean IoU (mIoU) and per-class IoU scores}.}\label{tab:strategy5-results}
\end{table*}

\paragraph*{Fine-grained segmentation with multiple layer}
This experiment implements Strategy 4 and 5 and provides results shown in Table \ref{tab:strategy4-results} and Table \ref{tab:strategy5-results}  respectively. Layer 1 behaves as in other experiments. In Strategy 4 the distinction between upper and lower garments allows a reduction of the involved labels for both Layer 2 and Layer 3. In this fashion, there is less ambiguity in choosing the labels that leads to an improvement of results. Strategy 5 exhibits similar behavior to Strategy 3 for fine-grained segmentation. Layer 2 provides very promising results for the visible part. Layer 3 provides satisfying results even if it is more challenging. Note that since in our dataset, the hidden part is always generated by the upper garment that occludes the lower garment the possible labels of layer 3 are chosen among only the lower garment types. 
Table \ref{fig:strategies} presents a visual comparison of the outcomes from various strategies applied to a sample of our synthetic dataset. The image features a complex shape with a long skirt, demonstrating how our approaches effectively identify not only the clothing segments but also the underlying body and overlapping regions with accuracy. 

\revision{Further experiments on our synthetic dataset are reported in Table \ref{fig:gim3d_results}. Several subjects with varying outfits have been evaluated for different strategies. Here we involved the Transformer architecture with data augmentation.
The underlying body is correctly estimated in all the examples. In the example of row $4$ the legs are hidden by the long skirt. The garment  parts are well highlighted also with fine-grained segmentation (rows $2-4$). The overlapping regions are detected with high accuracy. Note that, as expected, in the example of row $4$ no hidden segments are found on layer 2 since the garments are well separated by the belly that is correctly labeled as body part on layer 1.}

\paragraph*{Generalization to real data}
\revision{We evaluated our cloth layering segmentation method on real data of the CloSe Dataset \cite{antic2024close}. We provided  visual results only since the segmentation ground truth for each layer is unavailable. Table \ref{fig:close_result} shows results for each strategy of some subjects with various garments. The segmentation on these real samples is very satisfying despite the training has been carried out on synthetic data. The detected garments and body structures appear visually coherent for both coarse and fine-grained segmentation approaches. According to our cloth layering framework, the cloth overlap and underlying body are well estimated in each strategy providing reliable cloth layering outcomes also on these challenging real examples. }

\revision{
\subsection{Discussion}
\paragraph*{Backbones}
Among the evaluated architectures, Point Transformer consistently achieved the best performance in all implemented strategies. This outcome is consistent with the findings reported in~\cite{zhao2021point}, which demonstrated superior performance over previous state-of-the-art methods. The improved results can be attributed, in part, to the model’s ability to effectively handle a larger number of parameters; considering this aspect in all implemented architectures, the Point Transformer comprises approximately 7M parameters, compared to ~2M in DGCNN and ~1M in PointNet++.
}

\revision{
\paragraph*{Data augmentation}
As expected, data augmentation led to improved results for both Point Transformer and PointNet++, enhancing generalization and overall metric values.
However, in the case of DGCNN, data augmentation did not yield performance gains and in some cases even degraded the results.
This may be attributed to the model’s limited capacity to handle increased variability in the input, suggesting that DGCNN relies more heavily on local geometric features and is less robust to transformations such as rotation and translation~\textemdash~despite the presence of a spatial transformation network in its architecture~\cite{dgcnn}.
These observations further highlight the relative limitations of DGCNN in capturing global contextual information compared to more expressive models like Point Transformer.
}

\revision{
\paragraph*{Strategies}
As expected coarse segmentation (i.e., Strategy 1, 2, 3) performed better than fine-grained segmentation (i.e., Strategy 4, 5) since the latter is a much more challenging task. When explicit overlap is considered (i.e., Strategy 2 and 4) the obtained results are overall better than using implicit overlap (i.e., Strategy 3, 5). This can be explained by the less variability of the involved labels in the explicit overlap case. Moreover, we highlight that the worst performances are observed in Strategies 3 and 5 for hidden layer estimation, which is indeed the most challenging task. On the other hand, these strategies encoded more information and are the most suitable to be extended to further semantic cloth layers.   
}

\revision{
\paragraph*{Failure cases}
Table \ref{fig:failure cases} shows some failure cases, mainly derived from fine-grained segmentation. The first two examples are synthetic scans of our dataset, while the last two on the right are real scan data from \cite{antic2024close}. 
From the left, in the first example, we show Layer 2 from Strategy 5. The method correctly identifies the top, while it mislabels the trousers as shorts, in this case, the error is probably due to the length of the garment that is halfway between shorts and trousers. 
The second subject is in a very challenging pose (i.e., a jump).  In this case, we also see the prediction of Layer 2 from Strategy 5. The upper body is incorrectly labeled as top instead of long-shirt and in the lower body part, the network correctly identifies the shorts, but a small portion of them was mistakenly labeled as a t-shirt. 
In the third result, we show Layer 3 from Strategy 4. In this case, the skirt is correctly detected, and the prediction provides a good approximation of the part underneath the long-shirt. However, the method confuses the skirt label with the shorts label. 
In the last failure case, we see Layer 2 from Strategy 3. In this case, a challenging pose is also evaluated showing hands inside the pockets. Note that this kind of scenario was not included in our training dataset. Although the point cloud is well classified, with respect to the garment classes involved, some mistakes are highlighted where the hands are hidden in the pockets.
}

\begin{table*}[tbp]
\centering
\begin{tabular}{l l*{3}{>{\centering\arraybackslash}m{4cm}}}
\toprule
     & & Layer 1 & Layer 2 & Layer 3 \\
\midrule
\multirow{2}{*}{
  \parbox[c][5.4cm][c]{1.2cm}{\centering\rotatebox[origin=c]{90}{Strategy 2}}
} &
\rotatebox[origin=c]{90}{\textbf{GT}} &
\includegraphics[width=\linewidth]{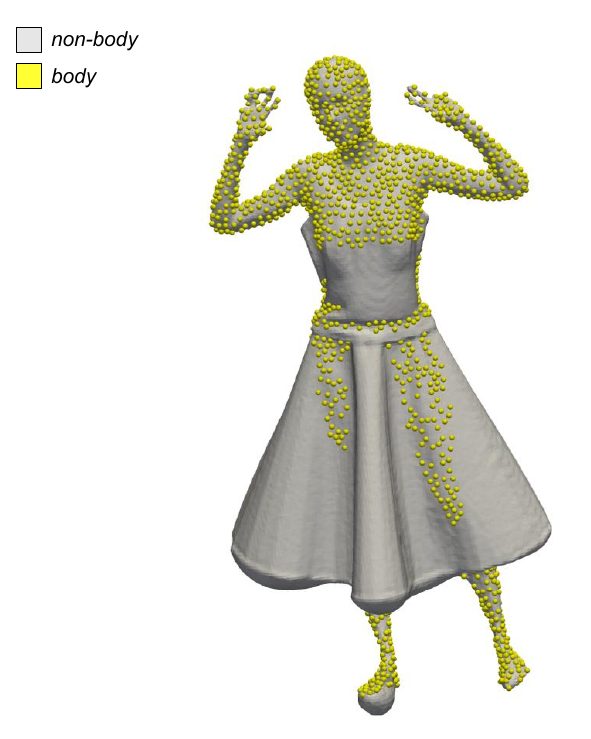} & 
\includegraphics[width=\linewidth]{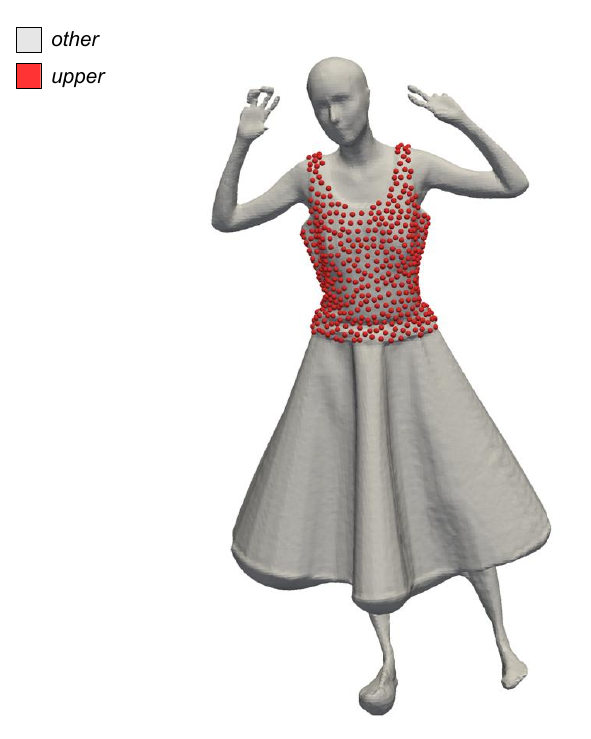} & 
\includegraphics[width=\linewidth]{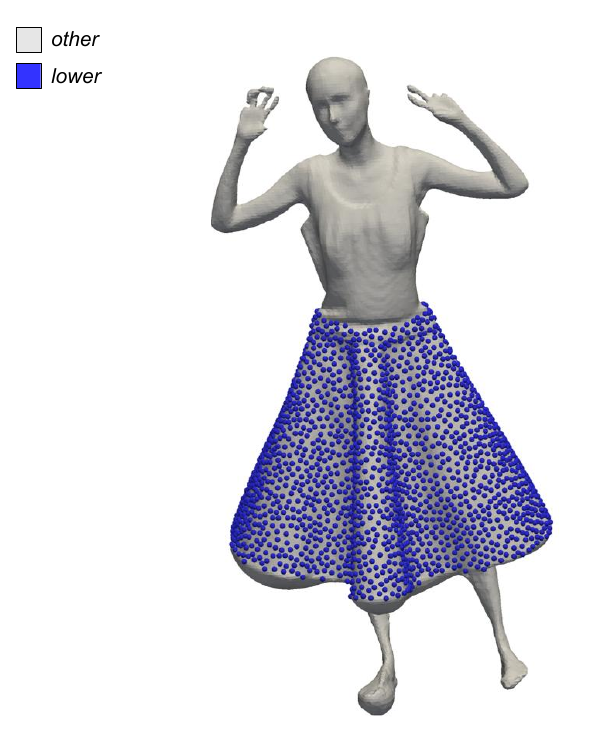} \\
& 
\rotatebox[origin=c]{90}{\textbf{PRED}} &
\includegraphics[width=\linewidth]{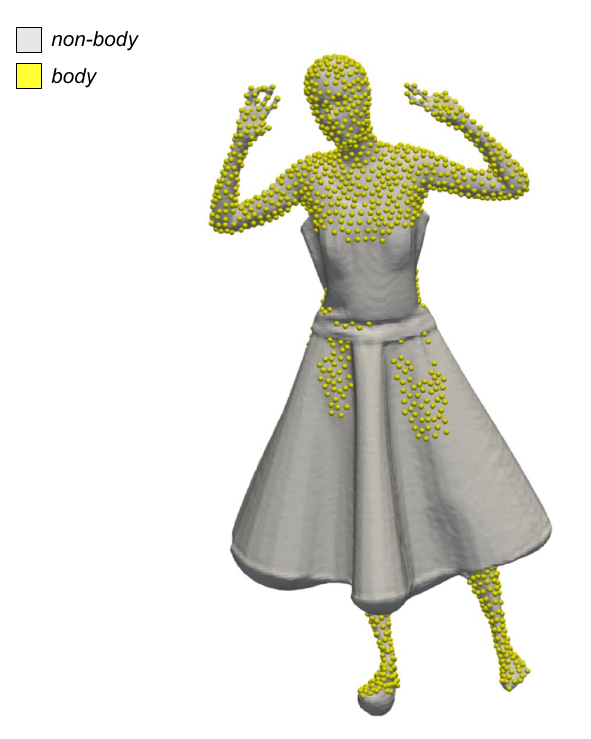}& 
\includegraphics[width=\linewidth]{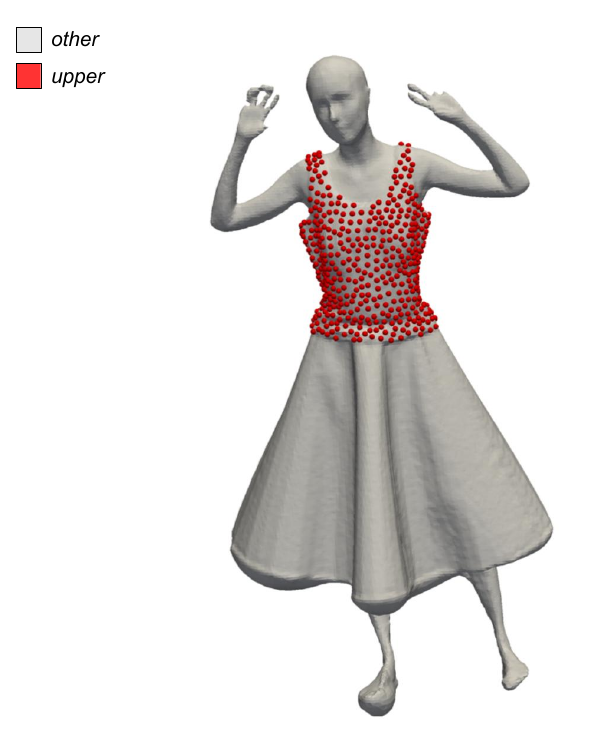}& 
\includegraphics[width=\linewidth]{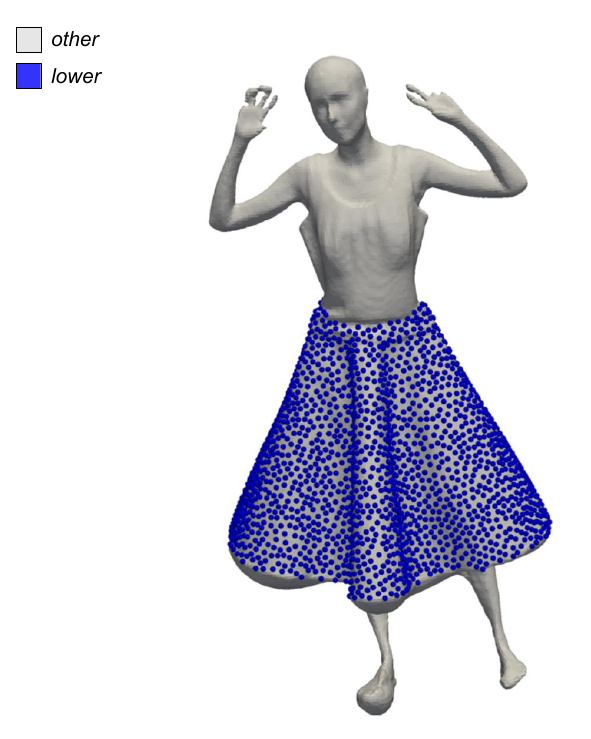}\\
\addlinespace
\bottomrule
\end{tabular}

\centering
\begin{tabular}{l l*{3}{>{\centering\arraybackslash}m{4cm}}}
\toprule
     & & Layer 1 & Layer 2 & Layer 3 \\
\midrule
\multirow{2}{*}{
  \parbox[c][5.4cm][c]{1.2cm}{\centering\rotatebox[origin=c]{90}{Strategy 3}}
} &
\rotatebox[origin=c]{90}{\textbf{GT}} &
\includegraphics[width=\linewidth]{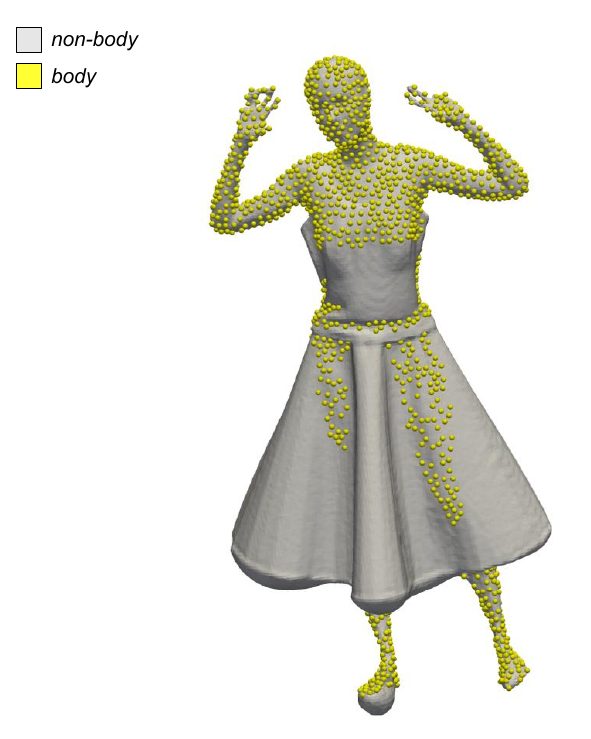} & 
\includegraphics[width=\linewidth]{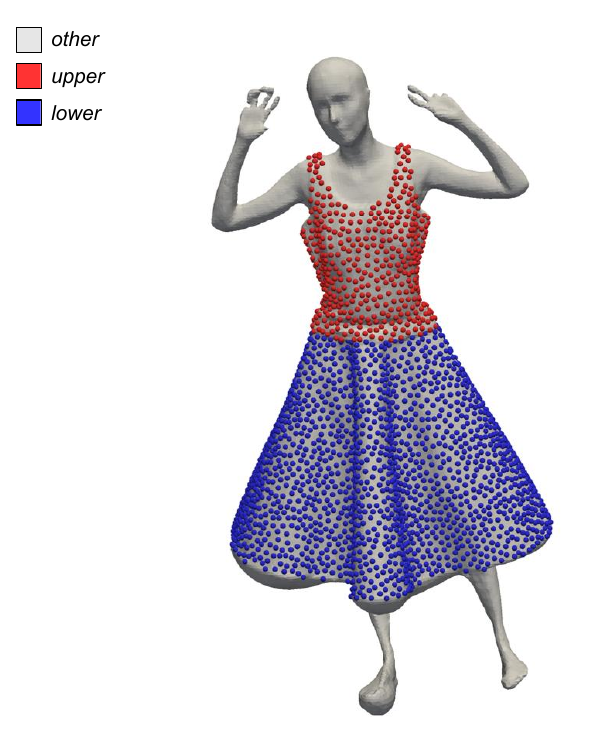} & 
\includegraphics[width=\linewidth]{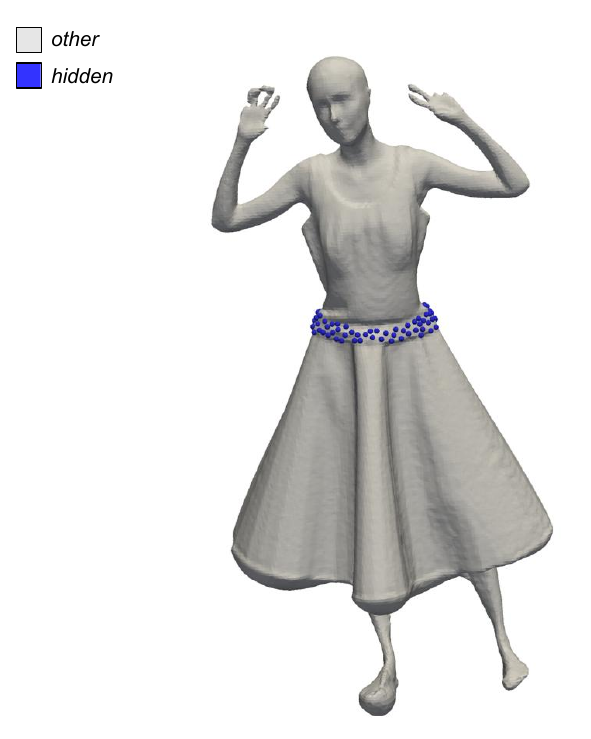} \\
& 
\rotatebox[origin=c]{90}{\textbf{PRED}} &
\includegraphics[width=\linewidth]{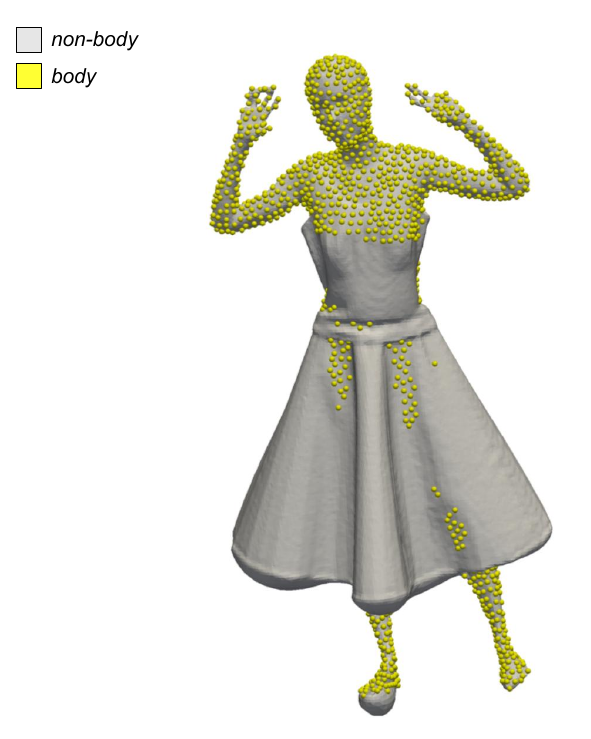}& 
\includegraphics[width=\linewidth]{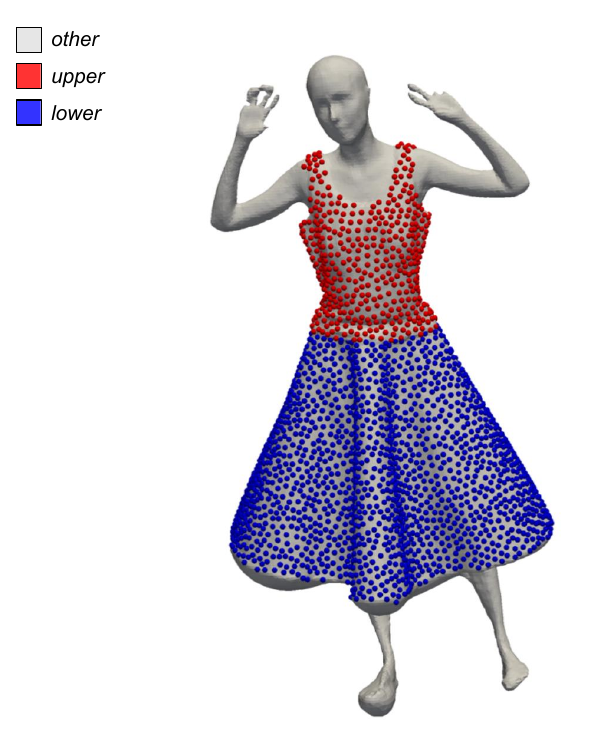}& 
\includegraphics[width=\linewidth]{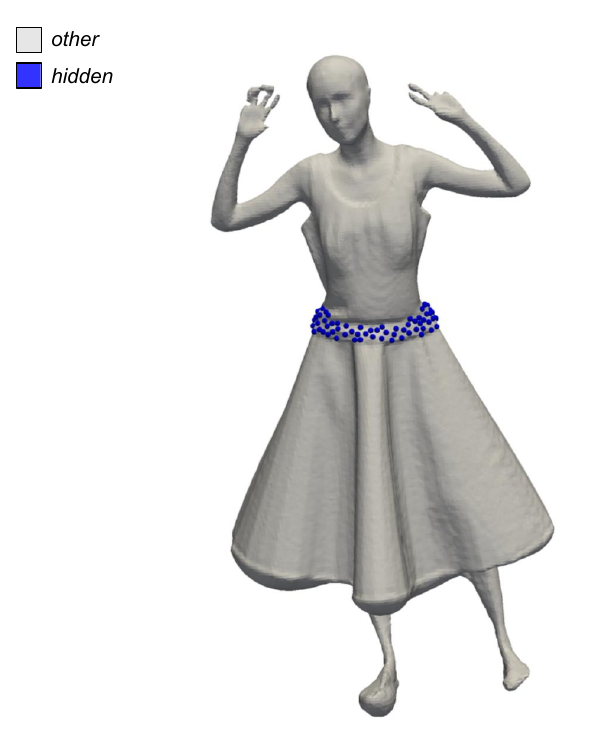}\\

\addlinespace
\bottomrule
\end{tabular}

\end{table*}

\begin{table*}[tbp]
\centering
\begin{tabular}{l l*{3}{>{\centering\arraybackslash}m{4cm}}}
\toprule
     & & Layer 1 & Layer 2 & Layer 3 \\
\midrule
\multirow{2}{*}{
  \parbox[c][5.4cm][c]{1.2cm}{\centering\rotatebox[origin=c]{90}{Strategy 4}}
} &
\rotatebox[origin=c]{90}{\textbf{GT}} &
\includegraphics[width=\linewidth]{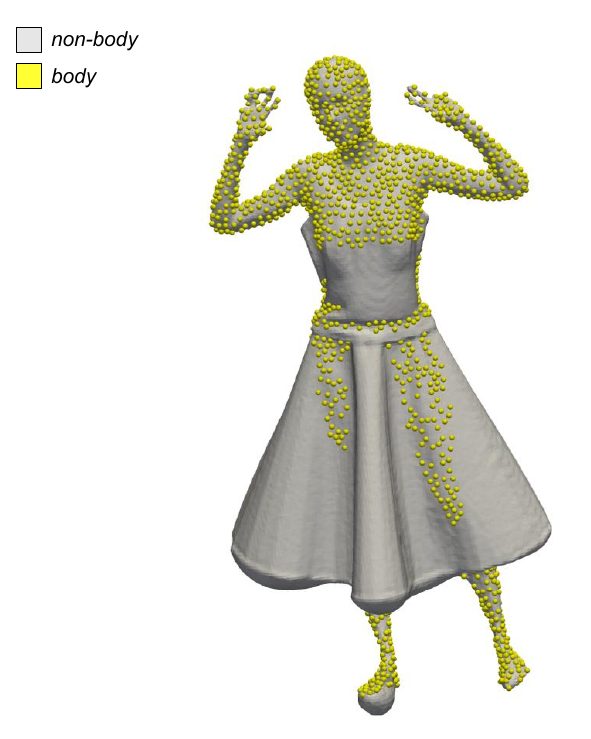} & 
\includegraphics[width=\linewidth]{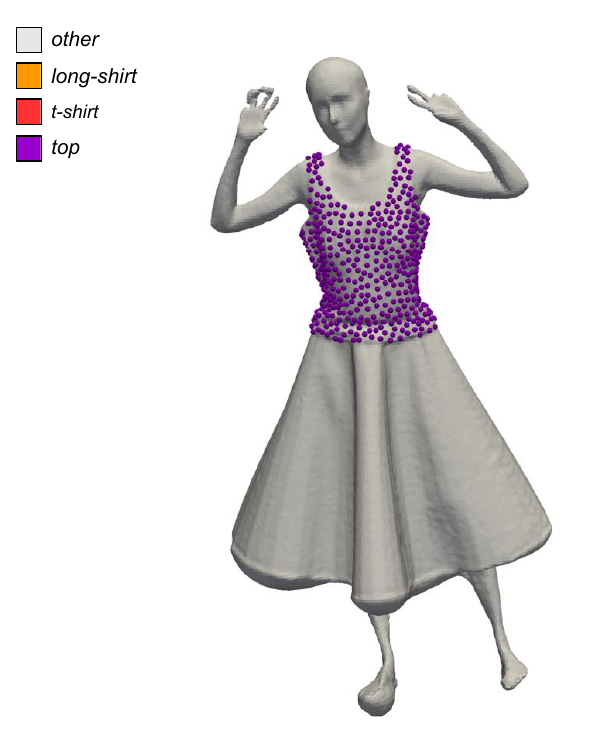} & 
\includegraphics[width=\linewidth]{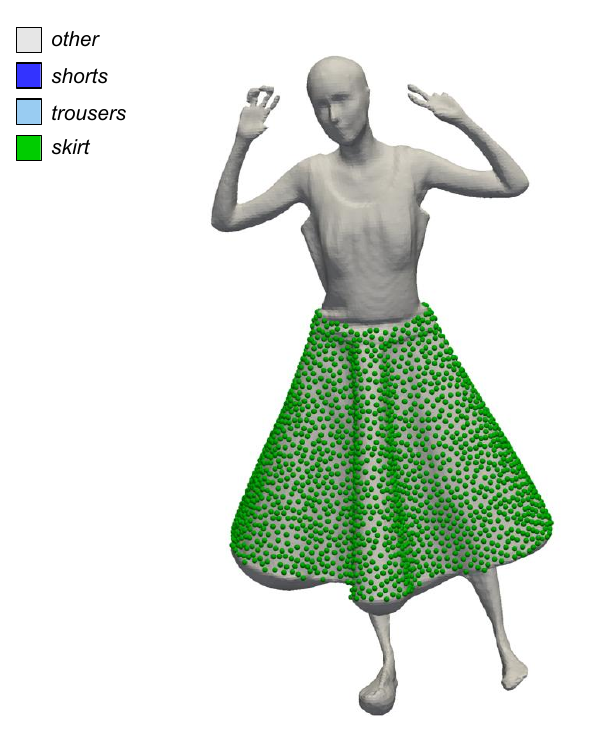} \\
& 
\rotatebox[origin=c]{90}{\textbf{PRED}} &
\includegraphics[width=\linewidth]{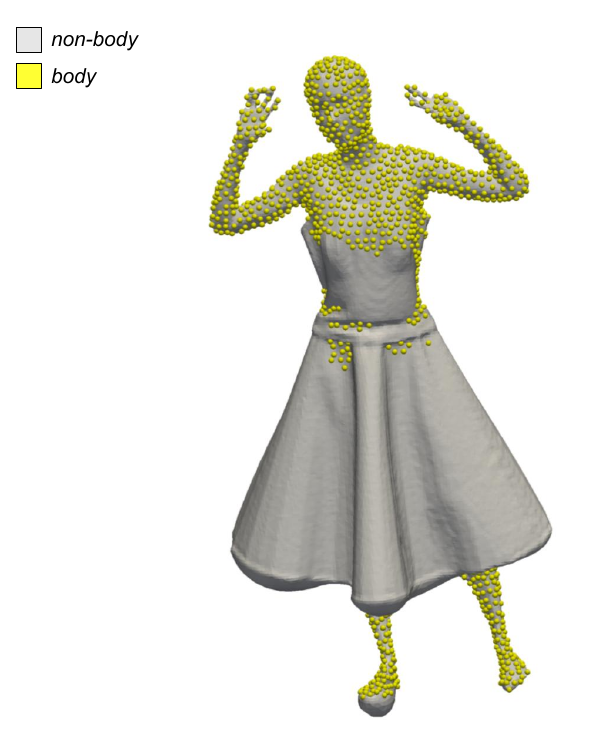}& 
\includegraphics[width=\linewidth]{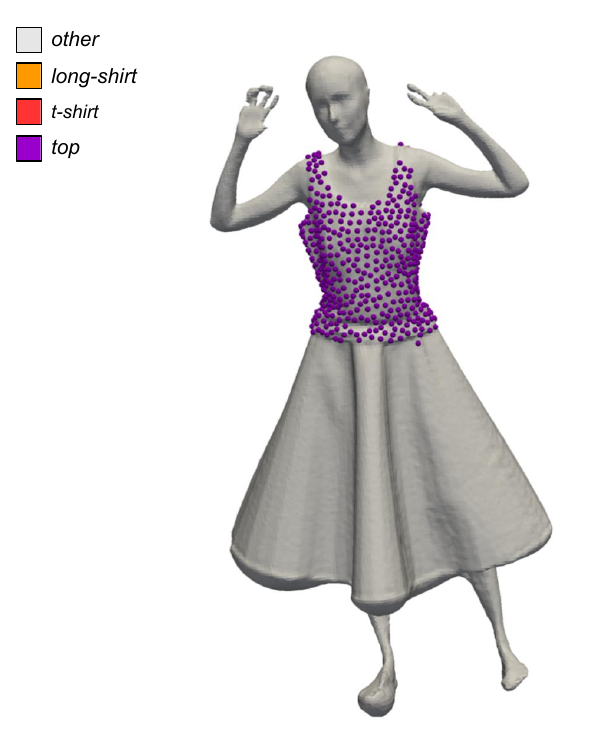}& 
\includegraphics[width=\linewidth]{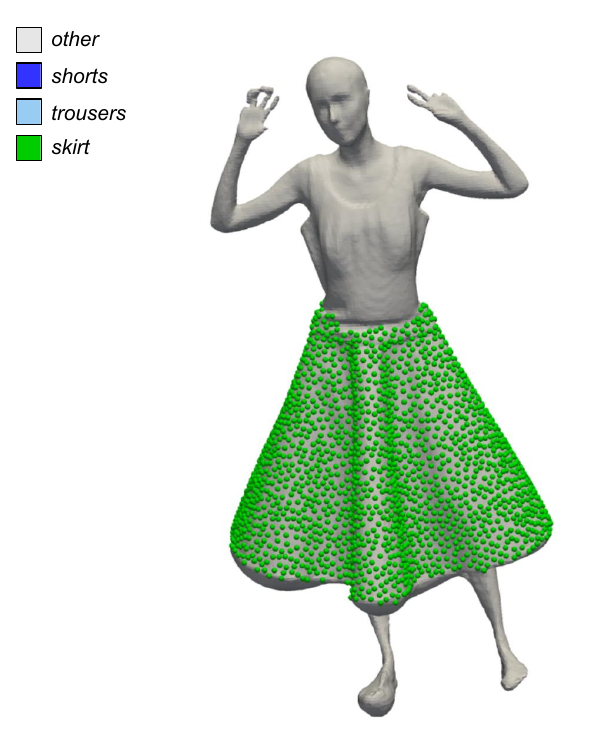}\\
\addlinespace
\bottomrule
\end{tabular}

\centering
\begin{tabular}{l l*{3}{>{\centering\arraybackslash}m{4cm}}}
\toprule
     & & Layer 1 & Layer 2 & Layer 3 \\
\midrule
\multirow{2}{*}{
  \parbox[c][5.4cm][c]{1.2cm}{\centering\rotatebox[origin=c]{90}{Strategy 5}}
} &
\rotatebox[origin=c]{90}{\textbf{GT}} &
\includegraphics[width=\linewidth]{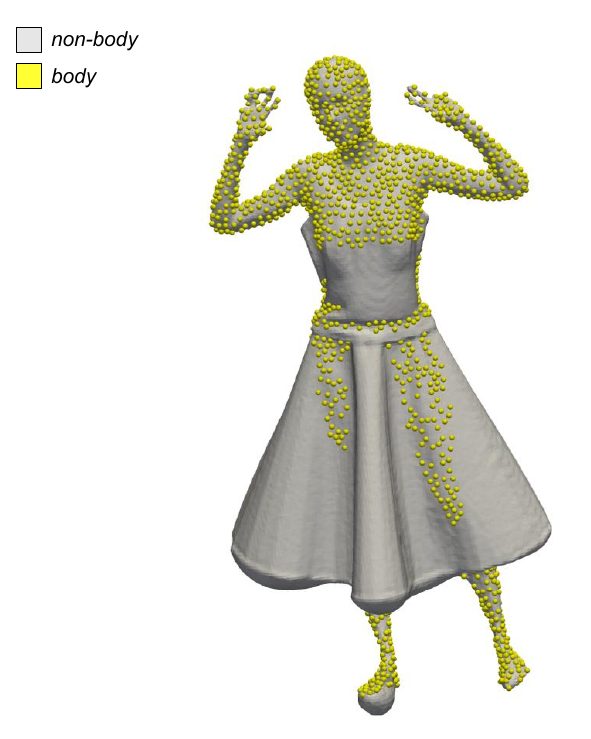} & 
\includegraphics[width=\linewidth]{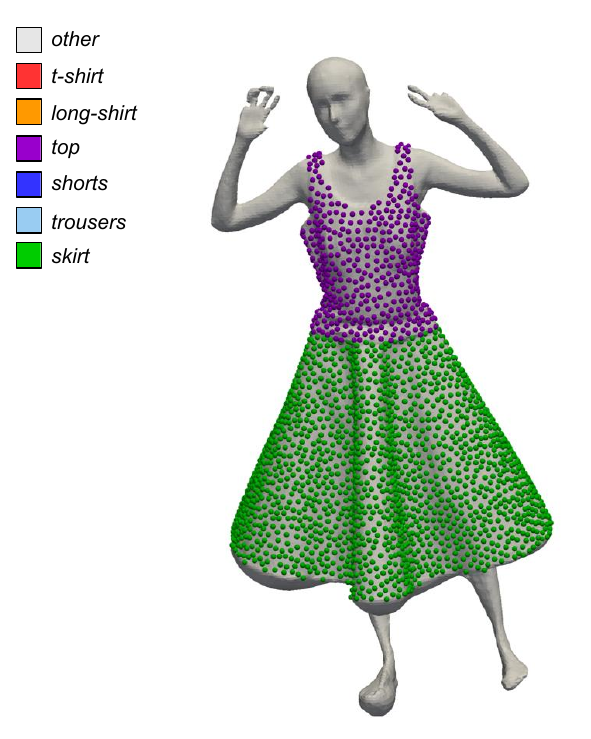} & 
\includegraphics[width=\linewidth]{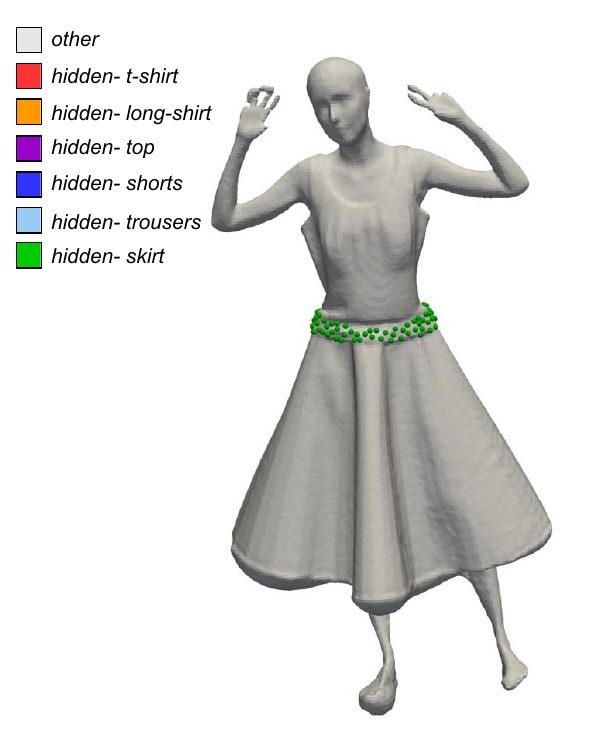} \\
& 
\rotatebox[origin=c]{90}{\textbf{PRED}} &
\includegraphics[width=\linewidth]{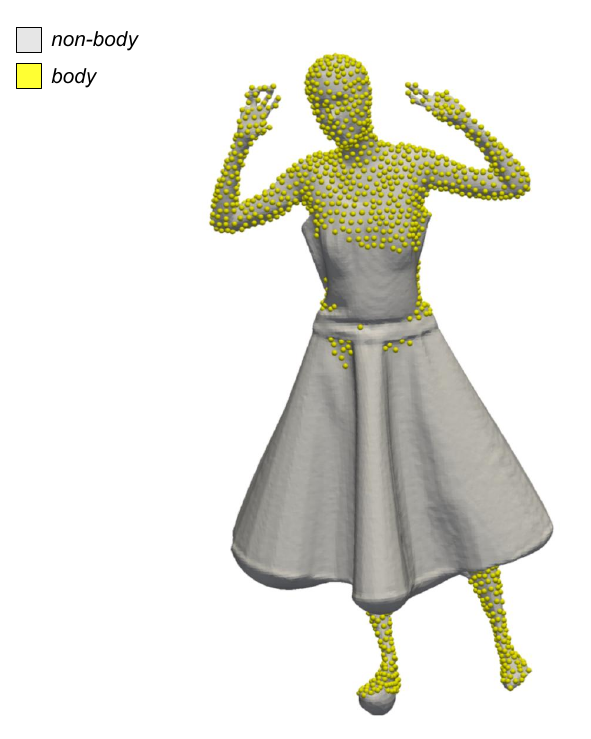}& 
\includegraphics[width=\linewidth]{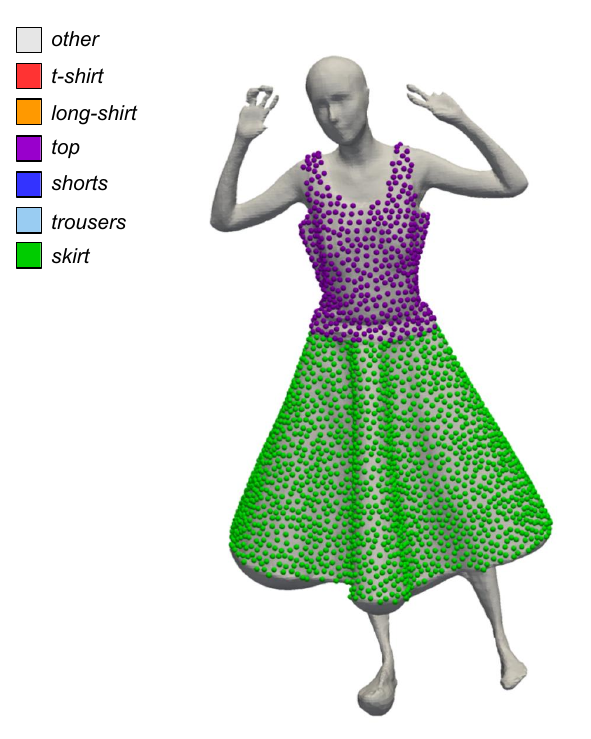}& 
\includegraphics[width=\linewidth]{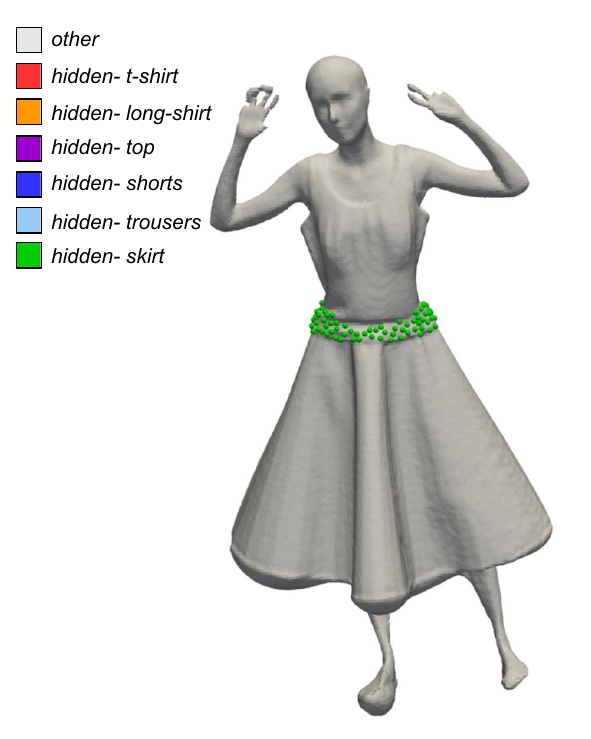}\\
\addlinespace
\bottomrule
\end{tabular}
\caption{
    \revision{Visualization of the results on our dataset, covering all evaluated strategies. Each table corresponds to a different strategy, with the first row showing the ground truth labels for each layer and the second row displaying the predictions produced by the network. Despite the variation in strategy, the results remain highly accurate, even for this particularly challenging case, which involves a long skirt.}
    }
\label{fig:strategies}
\end{table*}

\begin{table*}[tbp]
\centering
\begin{tabular}{l*{3}{>{\centering\arraybackslash}m{4cm}}}
\toprule
     &Layer 1 & Layer 2 & Layer 3 \\
\midrule
\rotatebox[origin=c]{90}{Strategy 3} &
\includegraphics[width=\linewidth]
{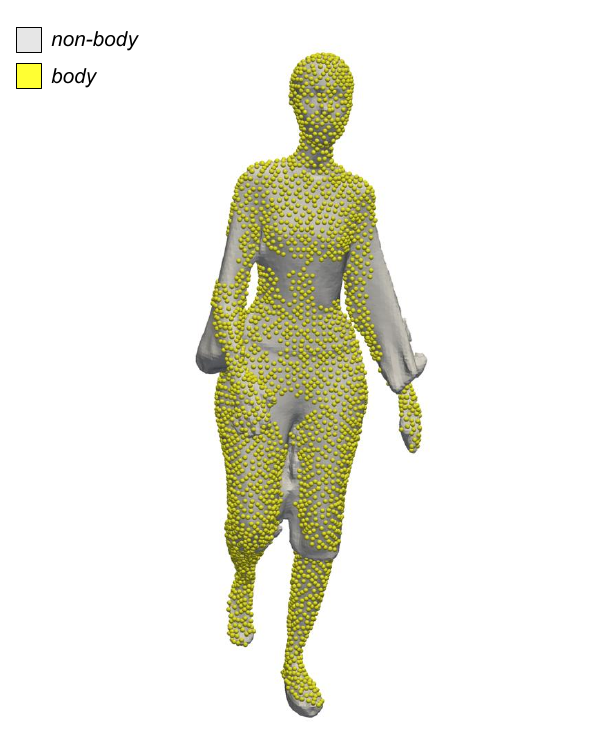} & 
\includegraphics[width=\linewidth]{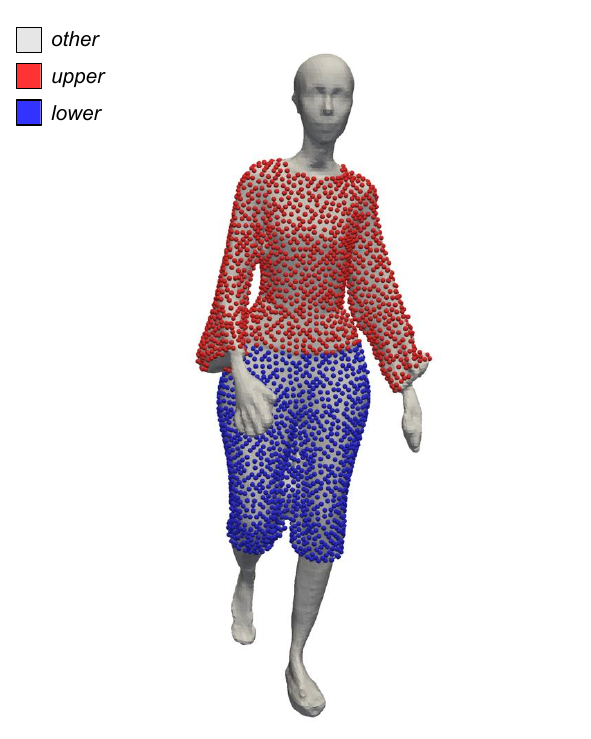} & 
\includegraphics[width=\linewidth]{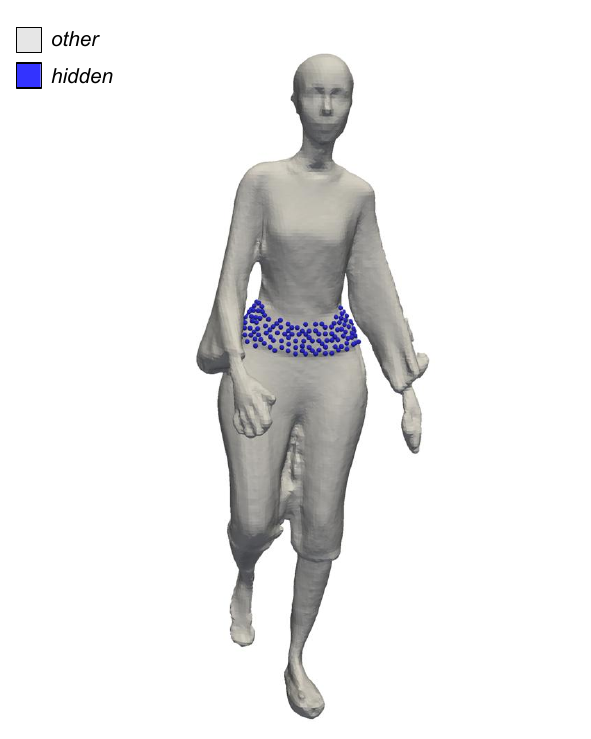} \\
\addlinespace
\rotatebox[origin=c]{90}{Strategy 4} &
\includegraphics[width=\linewidth]
{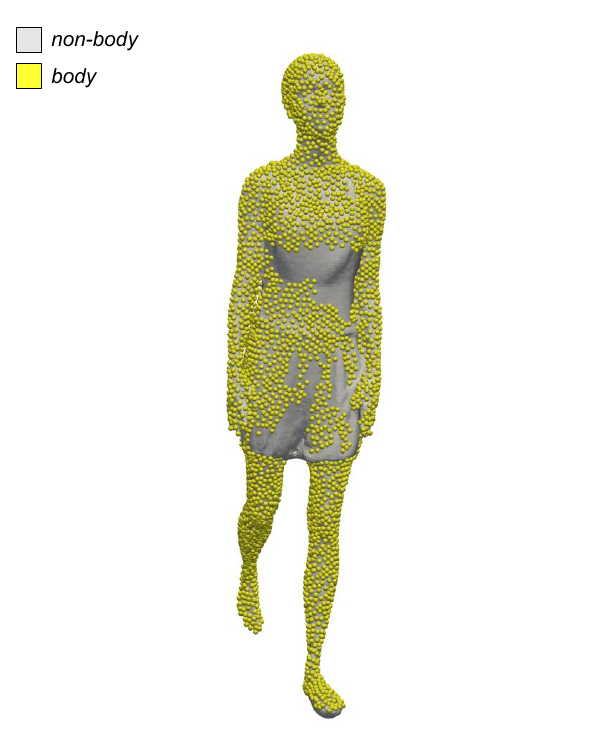} & 
\includegraphics[width=\linewidth]{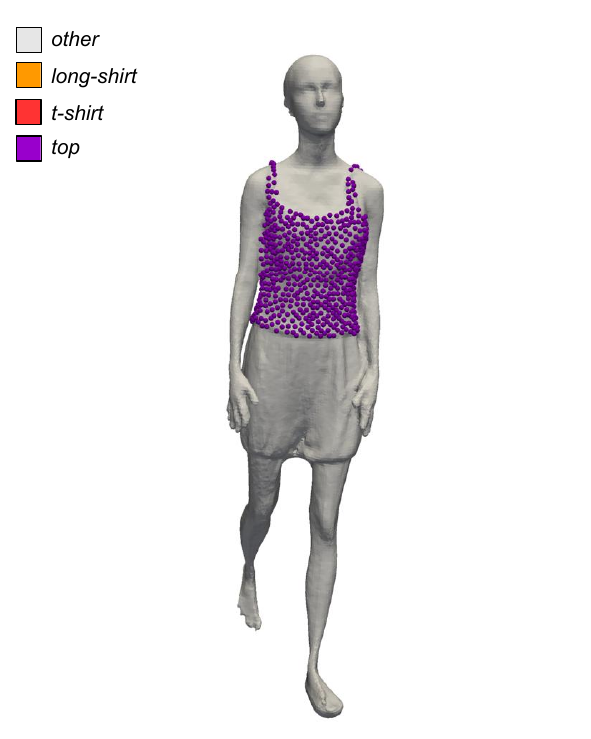} & 
\includegraphics[width=\linewidth]{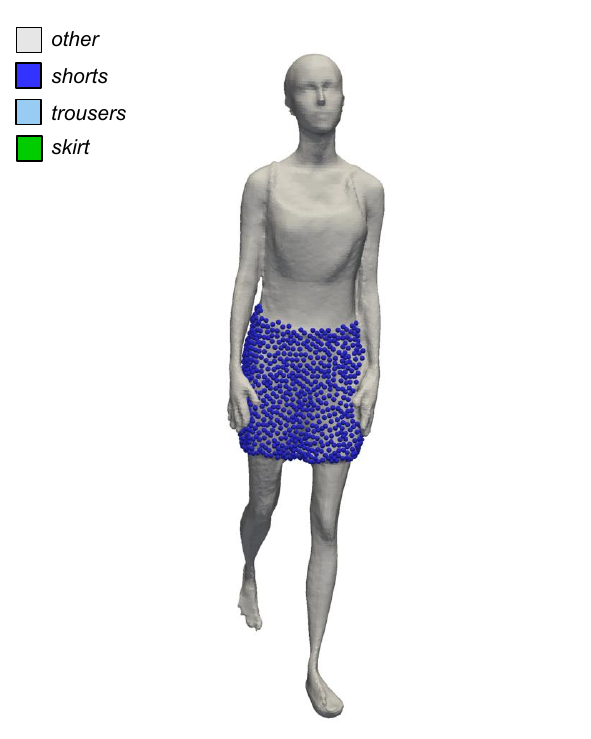} \\
\addlinespace
\rotatebox[origin=c]{90}{Strategy 5} &
\includegraphics[width=\linewidth]{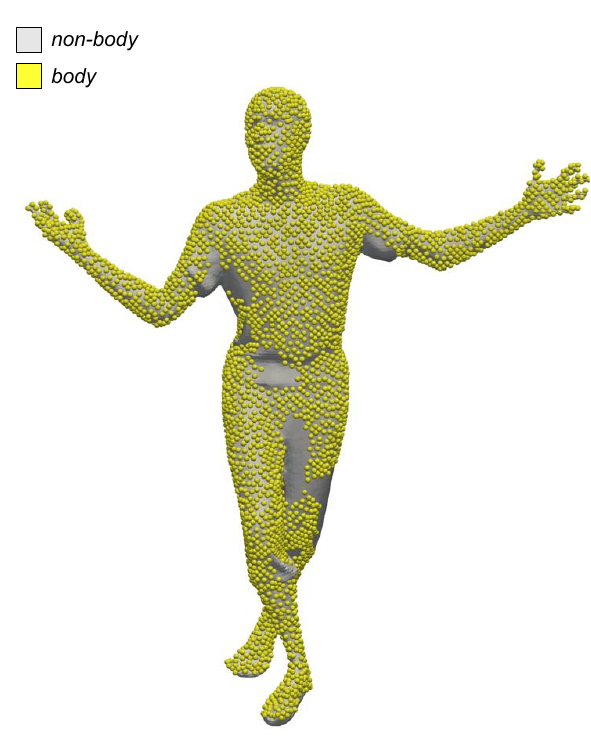} & 
\includegraphics[width=\linewidth]{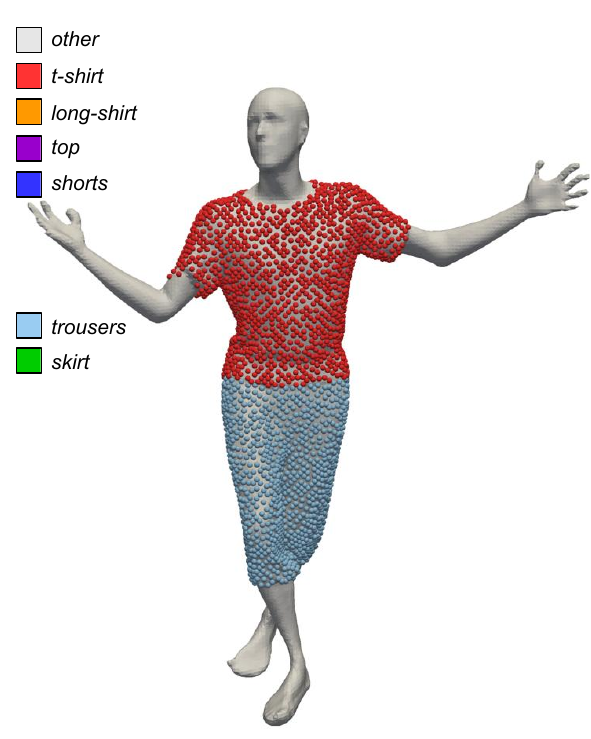} & 
\includegraphics[width=\linewidth]{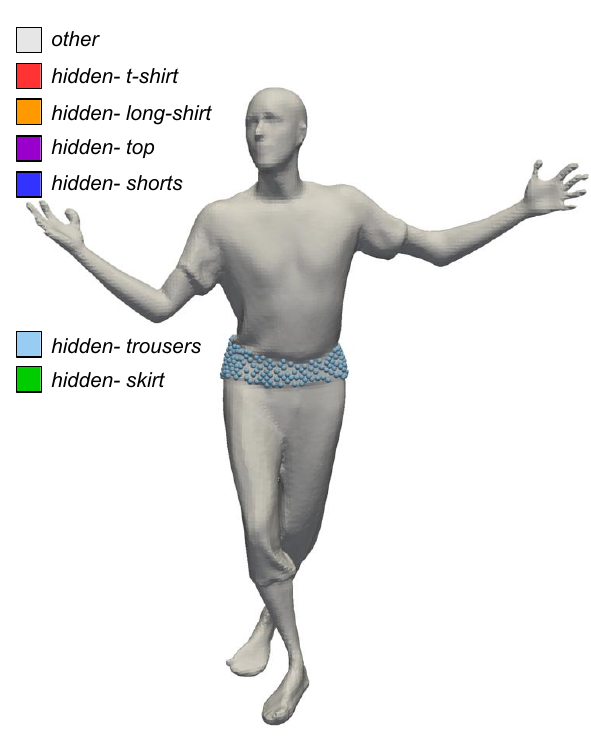} \\
\addlinespace
\rotatebox[origin=c]{90}{Strategy 5} &
\includegraphics[width=\linewidth]
{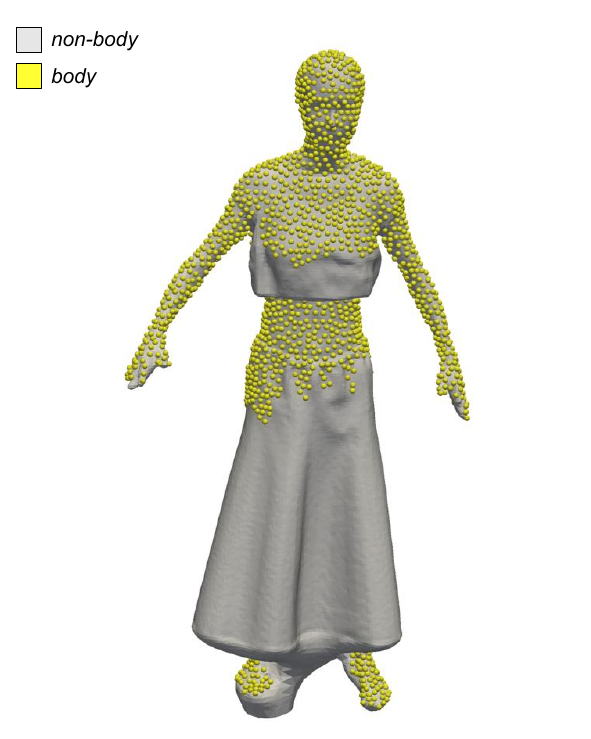} & 
\includegraphics[width=\linewidth]{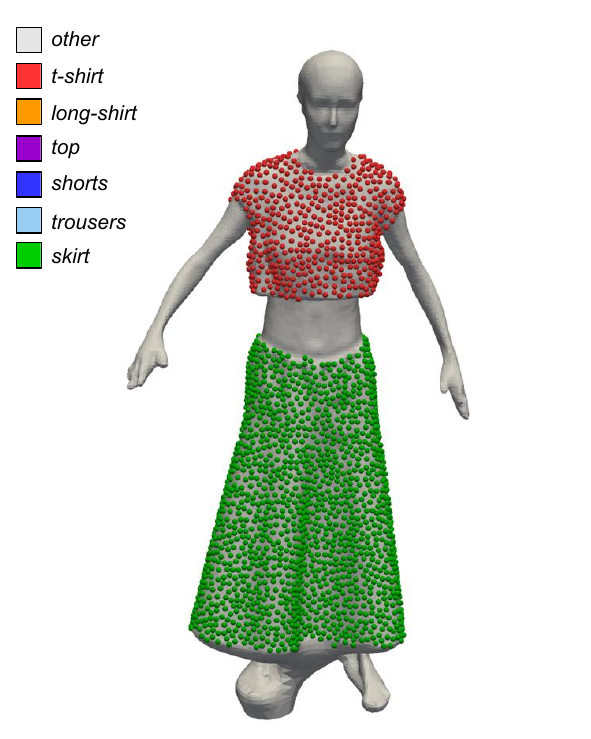} & 
\includegraphics[width=\linewidth]{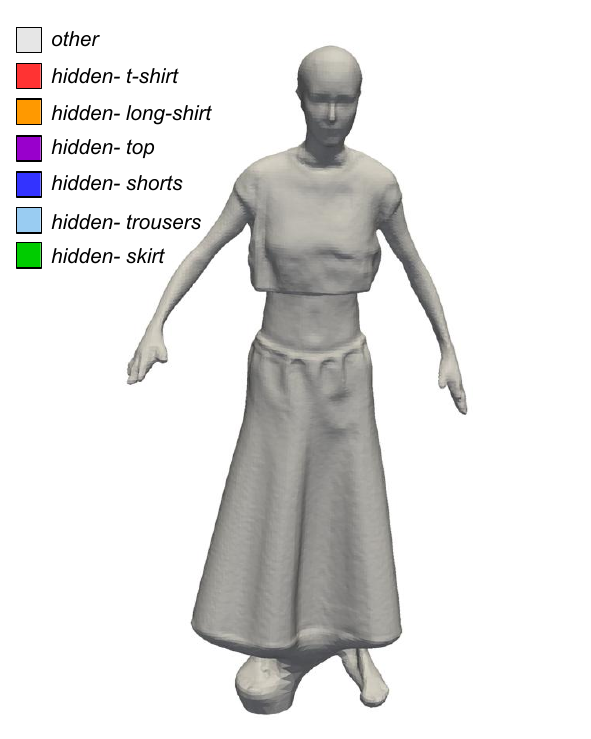} \\
\addlinespace
\bottomrule
\end{tabular}
\caption{
    \revision{Here we show some results performed on subjects from our dataset. We want to highlight the result in the third row that shows a highly accurate segmentation on a challenging pose (where the pants are merged by the scanner), while the result in the last row shows that the predicted labels are correct even in the absence of overlap. Furthermore, the estimation of the underlying body in layer 1 and the prediction of the skirt in layer 2 (that is a particularly challenging garment class in cloth segmentation tasks) demonstrate a high level of accuracy.}
    }
\label{fig:gim3d_results}
\end{table*}

\begin{table*}[tbp]
\centering
\begin{tabular}{l*{4}{>{\centering\arraybackslash}m{4cm}}}
\toprule
    & Input & Layer 1 & Layer 2 & Layer 3 \\
\midrule
\rotatebox[origin=c]{90}{Strategy 2} &
\includegraphics[width=\linewidth]{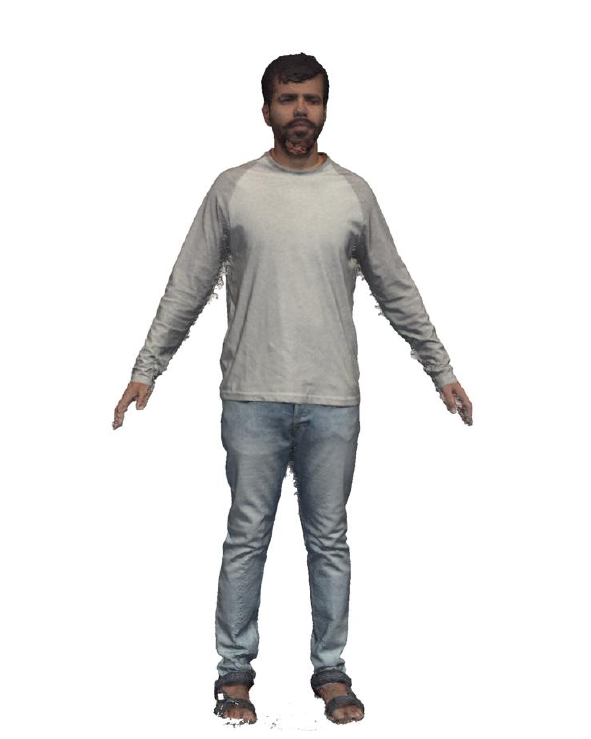} & 
\includegraphics[width=\linewidth]{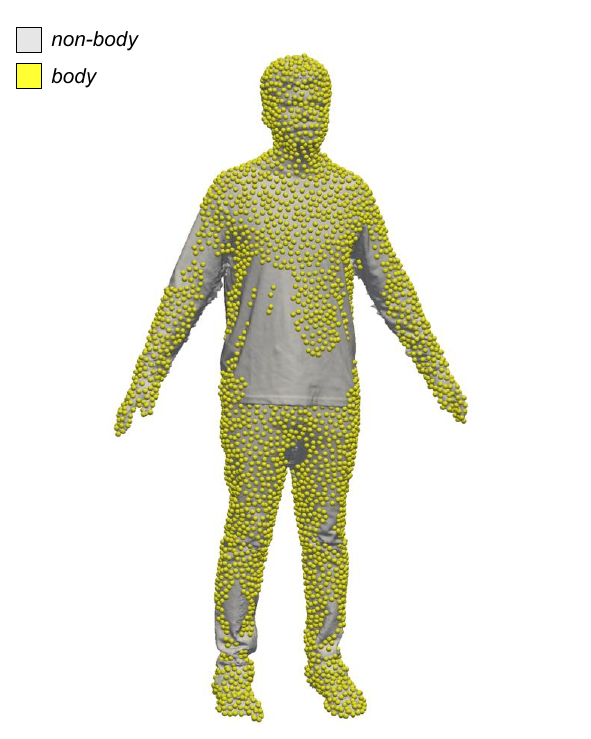} & 
\includegraphics[width=\linewidth]{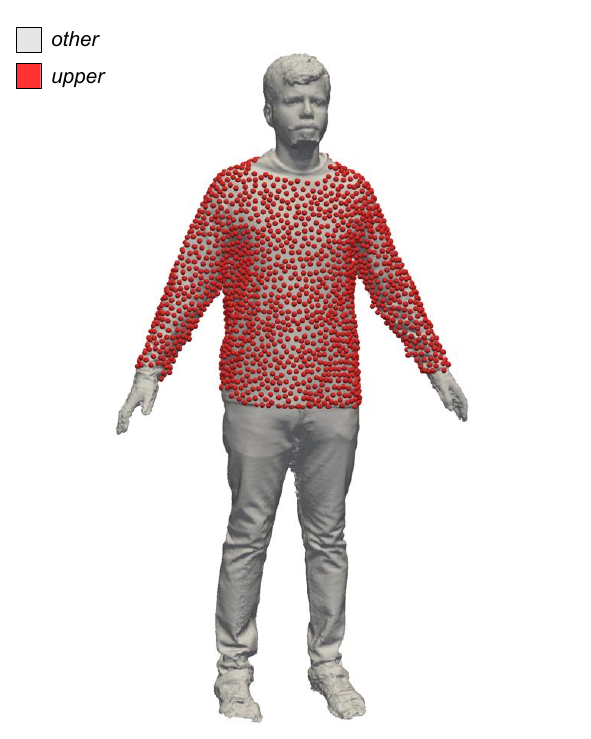} & 
\includegraphics[width=\linewidth]{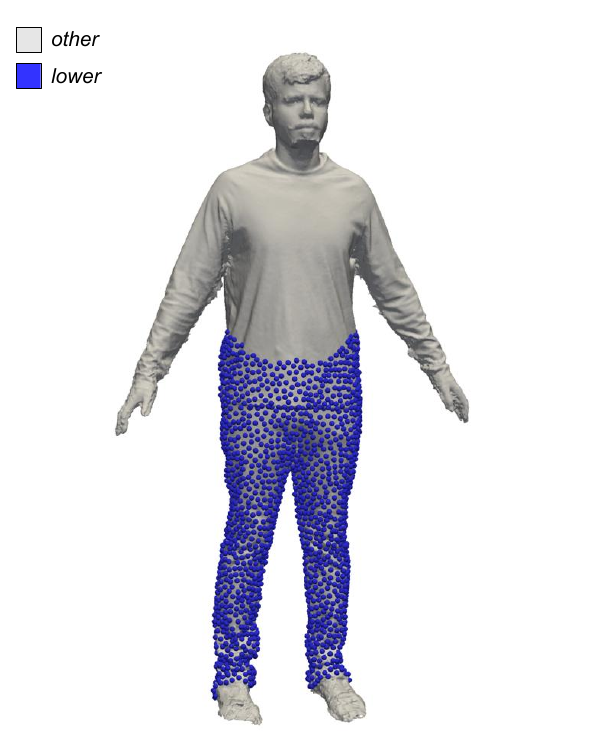} \\
\addlinespace
\rotatebox[origin=c]{90}{Strategy 3} &
\includegraphics[width=\linewidth]{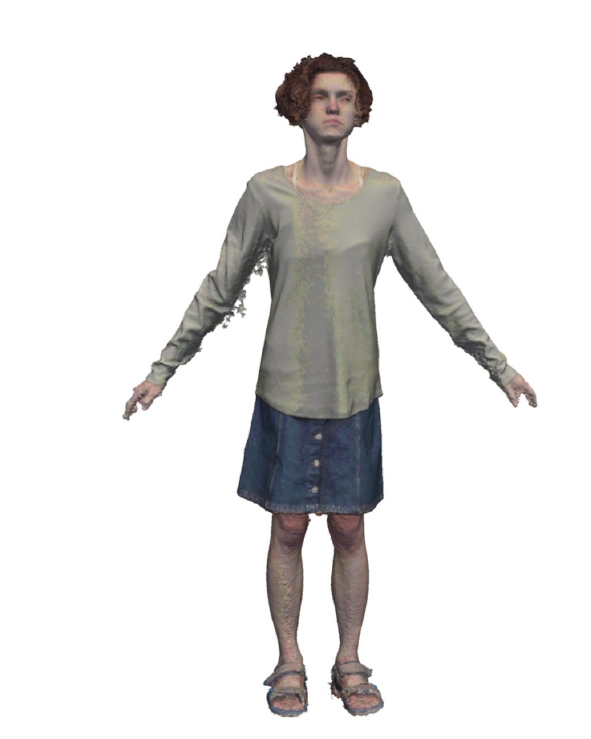} & 
\includegraphics[width=\linewidth]{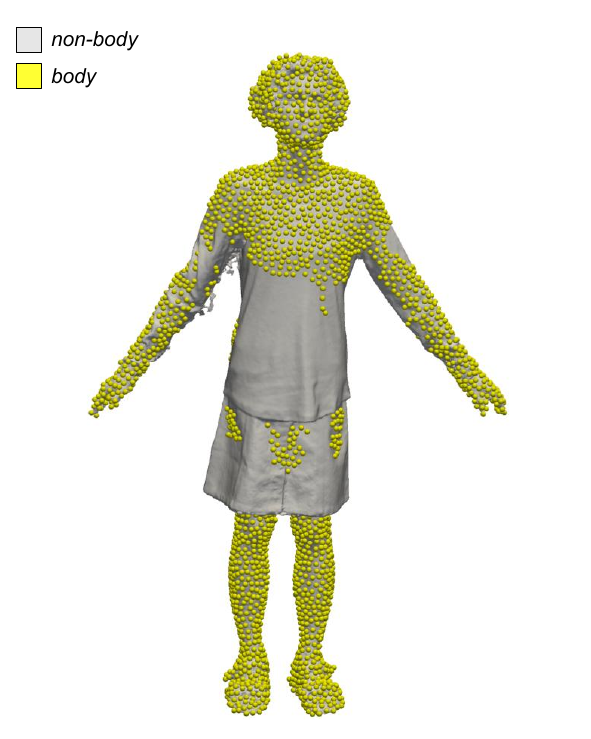} & 
\includegraphics[width=\linewidth]{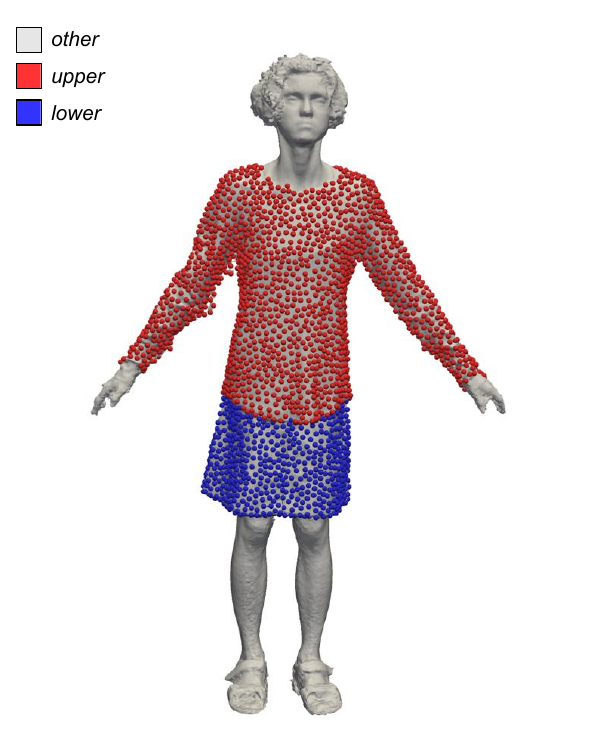} & 
\includegraphics[width=\linewidth]{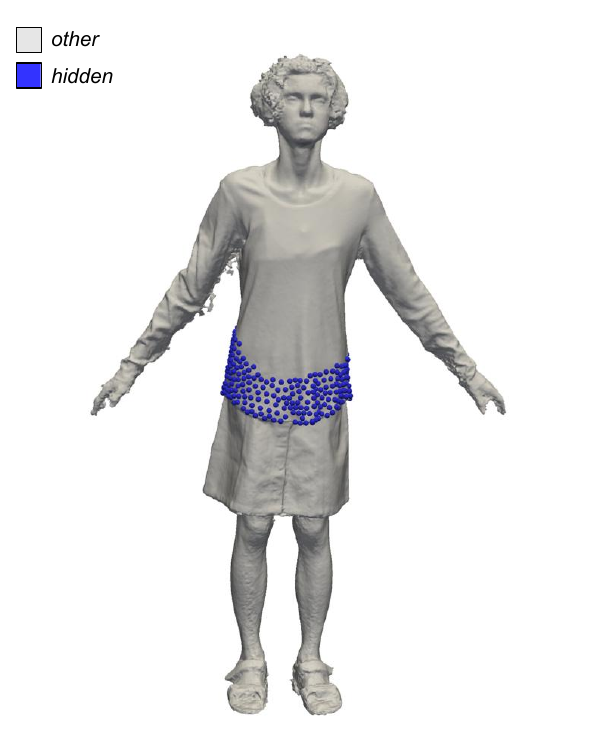} \\
\addlinespace
\rotatebox[origin=c]{90}{Strategy 4} &
\includegraphics[width=\linewidth]{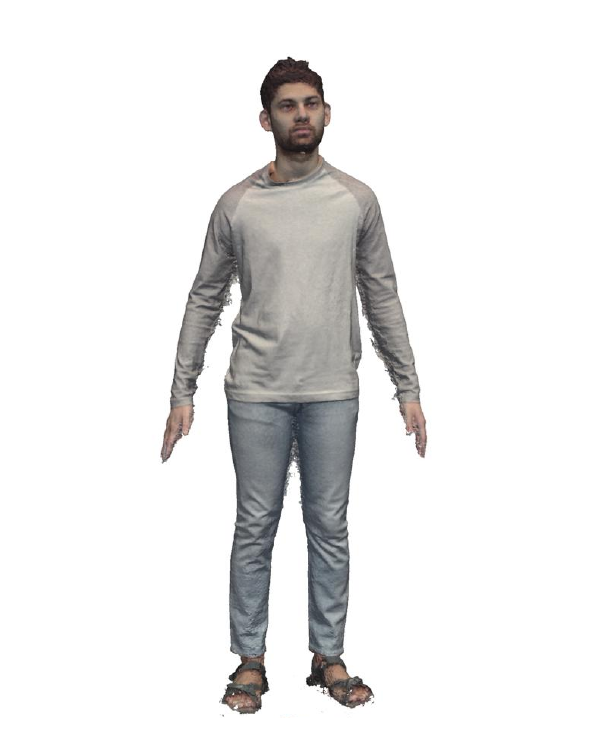} & 
\includegraphics[width=\linewidth]{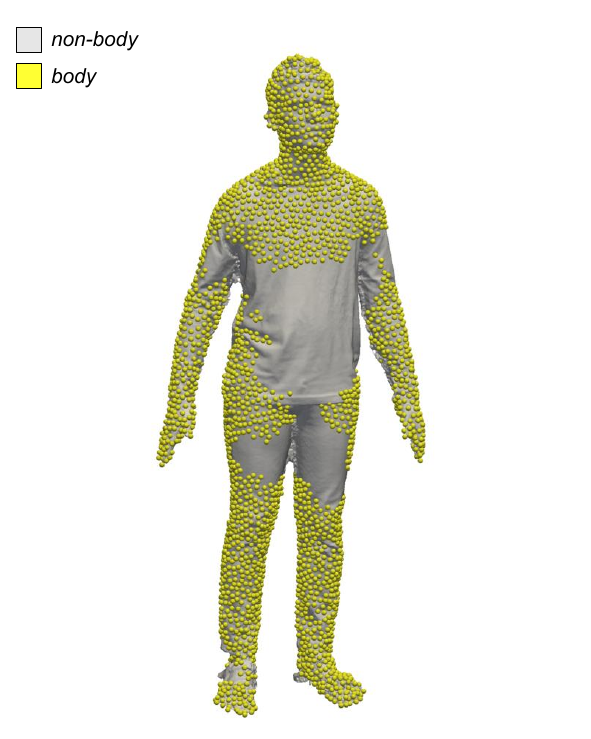} & 
\includegraphics[width=\linewidth]{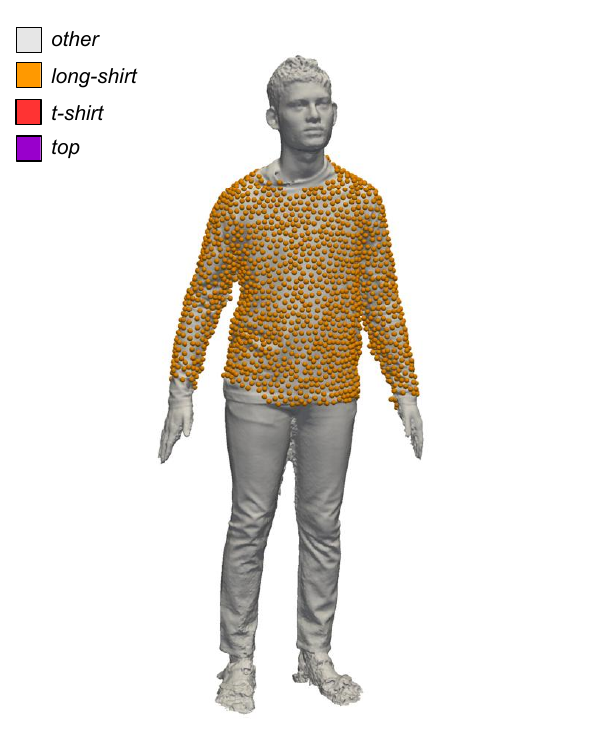} & 
\includegraphics[width=\linewidth]{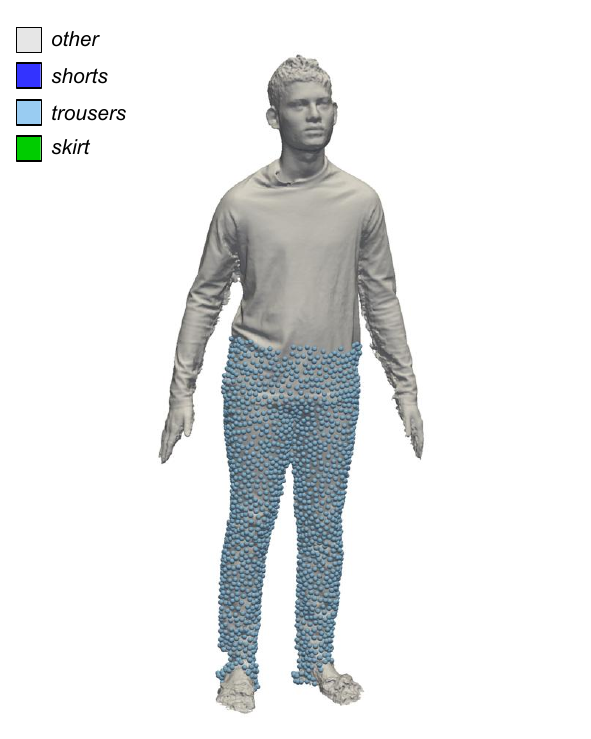} \\
\addlinespace
\rotatebox[origin=c]{90}{Strategy 5} &
\includegraphics[width=\linewidth]{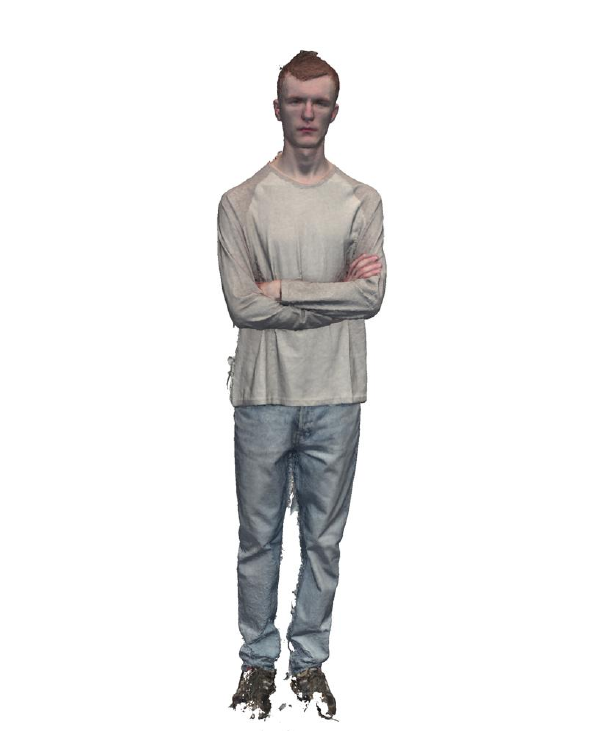} & 
\includegraphics[width=\linewidth]{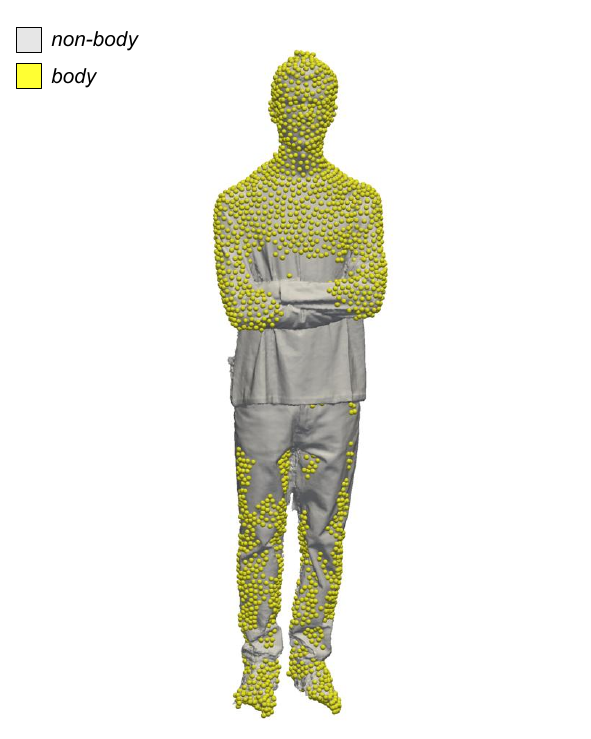} & 
\includegraphics[width=\linewidth]{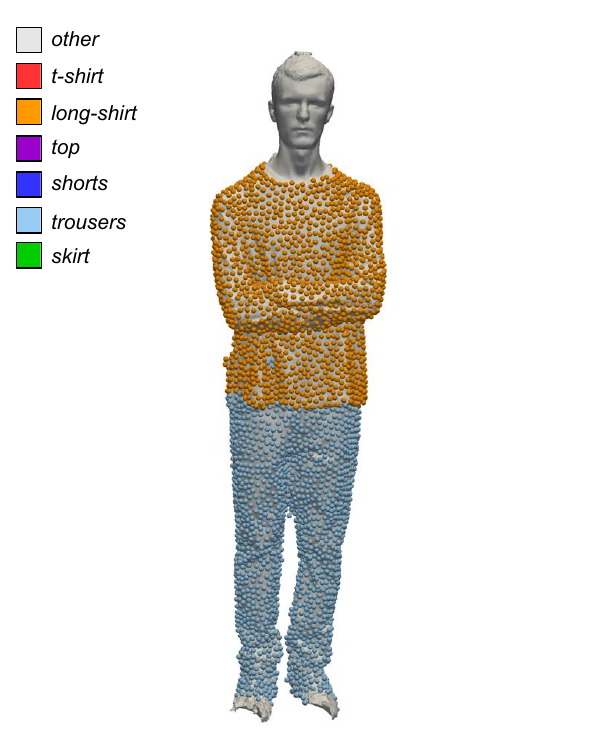} & 
\includegraphics[width=\linewidth]{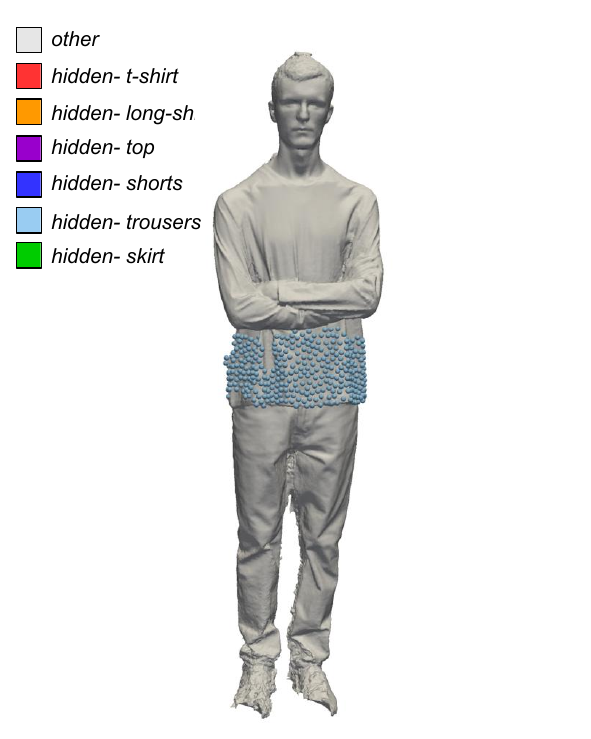} \\
\addlinespace
\bottomrule
\end{tabular}
\caption{
    \revision{Here some examples of real-scan results obtained using our multilayer segmentation method. These results demonstrates the ability of our method to handle variability in both subjects and their outfits. The segmentation accurately predicts the layer labels even in challenging cases, such as the second row featuring a skirt, and the last row, which shows a difficult pose with crossed arms. The predicted overlap also appears visually consistent, despite the lack of ground truth for quantitative evaluation, with the input shown in the first column was available.}
    }
\label{fig:close_result}
\end{table*}

\begin{table*}[tbp]
\centering
\begin{tabular}{*{4}{>{\centering\arraybackslash}m{4cm}}}
\toprule
     Strategy 5 &Strategy 5 & Strategy 4 & Strategy 3 \\
\midrule
\begin{overpic}[width=\linewidth]{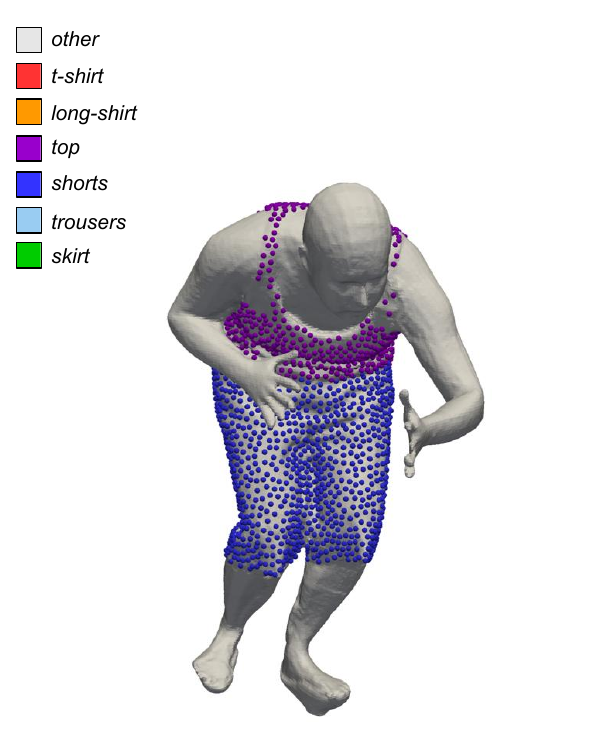} \put(29,-8){\color{black} Layer 2}
\end{overpic} &
\begin{overpic}[width=\linewidth]{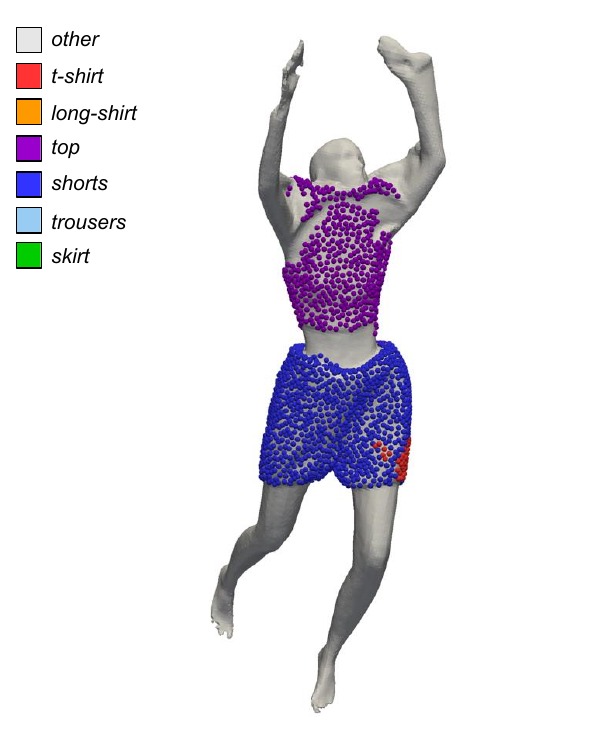} \put(30,-8){\color{black} Layer 2}
\end{overpic} & 
\begin{overpic}[width=\linewidth]{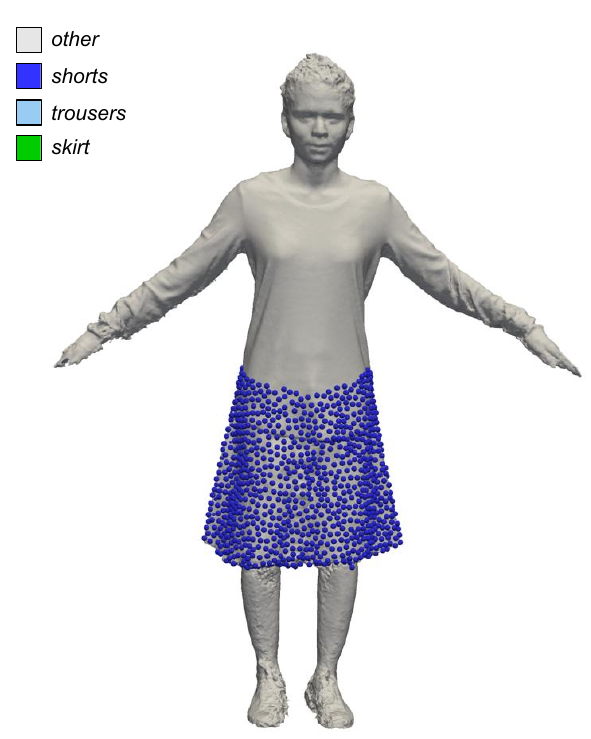} \put(30,-8){\color{black} Layer 3}
\end{overpic} &  
\begin{overpic}[width=\linewidth]{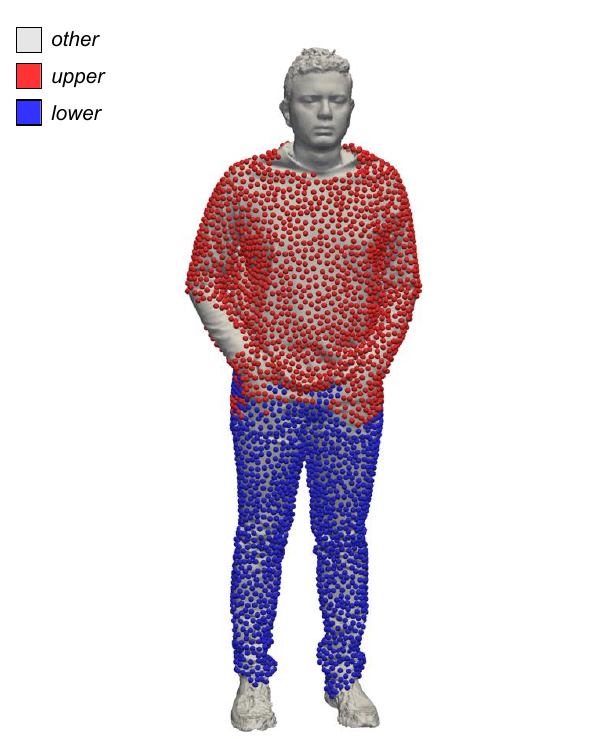} \put(30,-8){\color{black} Layer 2}
\end{overpic} \\
\addlinespace
\addlinespace
\addlinespace
\bottomrule
\end{tabular}
\caption{
    \revision{This figure presents four failure cases—two dataset scans and two real scans. In the first, a Layer 2 prediction from Strategy 5 correctly labels the top but misclassifies trousers as shorts. The second also uses Layer 2 from Strategy 5 and involves a difficult jumping pose, where the long-shirt is mislabeled as a top; the shorts are mostly correct, with a small section labeled as a t-shirt. The third case, from Layer 3 using Strategy 4, shows a well-segmented skirt and a coherent approximation under the long-shirt, but the skirt is mislabeled as shorts. In the final case, a Layer 2 prediction from Strategy 3 tackles a challenging pose with hands in pockets—a case that is absent from training data—and largely succeeds, though it mislabels around the pocket region.}
    }
\label{fig:failure cases}
\end{table*}

\section{Conclusions}
\label{sec:conclusions}
In this work, we have proposed clothed human layering as a new 3D segmentation approach that has been properly designed for the semantic reconstruction of point clouds acquired by a 3D scanning procedure for garment modeling purposes. The overall aim is the automatic identification of non-disjoint point subsets that can be converted in the semantic human clothed layers namely the underlying body and the involved garment types. The key idea consists of allowing the detection of garment overlap by exploiting and evaluating different solution strategies. We have created a new dataset that simulates a real 3D sensor by generating very realistic 3D point clouds. Our approach exploits only geometric properties even if its extension to colored subjects is trivial. Our experimental results showed promising performances on the new dataset in all the evaluated strategies. Moreover, we have highlighted the capability of our synthetic dataset in generalizing the performances to challenging real-world examples. 

\paragraph*{Future work}
 There are several directions to be explored to improve our work. From the methodological aspects, it is possible to better exploit the interrelation between the involved layers and labels. Moreover, the segmentation procedure should be integrated with outfit segmentation to guide the selection of local labels from the global choice. 
 \revision{
 The idea of cloth layering can be further extended to other layers such as the case of a jacket over a t-shirt over a skirt. These supplemental clothing layers introduce complexity that cannot be handled by our work at this level, but we have observed that Strategy $3$ and $5$ are the most suitable to support further clothing layers and we will investigate more in this extension in future works.  
 Also, the expansion to further garment classes can be addressed by adding new samples in our datasets to enlarge the training procedure accordingly. \minor{The dataset could be further extended by incorporating more complex poses and occlusions to better reflect the variety of real-world scans. Additionally, as discussed in the strategy section, another valuable enhancement would be the inclusion of new combinations of visible and non-visible garments. Currently, the dataset only includes samples where, in cases of overlapping garments, the lower garment is worn underneath. Introducing the opposite scenario—where the lower garment is on top—would require adjustments to the strategy settings, with the exception of Strategy 5, which explicitly encodes in the label which garment is visible and which is not. Implementing these extensions would significantly improve the dataset’s applicability to multilayer segmentation tasks.}
 Another important limitation that has been highlighted in this work is that the size of the concurring classes is unbalanced (e.g. body vs garments). In future work, we plan to exploit additional loss functions properly designed to address this issue. } 






\bibliographystyle{elsarticle-num-names} 
\bibliography{egbibsample}     

\end{document}